\pdfoutput=1

\documentclass[12pt,preprint]{aastex}

\shorttitle{Periodic LINEAR variables}
\shortauthors{Palaversa et al.}

\begin{document}

\title{Exploring the Variable Sky with LINEAR. III. Classification of Periodic Light Curves}

\author{
Lovro Palaversa\altaffilmark{\ref{Geneve}},
\email{lovro.palaversa@unige.ch}
\v{Z}eljko Ivezi\'{c}\altaffilmark{\ref{Washington},\ref{PMF},\ref{OH}},
Laurent Eyer\altaffilmark{\ref{Geneve}},
Domagoj Ru\v{z}djak\altaffilmark{\ref{OH}},
Davor Sudar\altaffilmark{\ref{OH}},
Mario Galin\altaffilmark{\ref{Geo}},
Andrea Kroflin\altaffilmark{\ref{PMF}},
Martina Mesari\'{c}\altaffilmark{\ref{PMF}}, 
Petra Munk\altaffilmark{\ref{PMF}},
Dijana Vrbanec\altaffilmark{\ref{PMF}},
Hrvoje Bo\v{z}i\'{c}\altaffilmark{\ref{OH}},
Sarah Loebman\altaffilmark{\ref{Washington}},
Branimir Sesar\altaffilmark{\ref{Caltech}},
Lorenzo Rimoldini\altaffilmark{\ref{Geneve},\ref{ISDC}},
Nicholas Hunt-Walker\altaffilmark{\ref{Washington}},
Jacob VanderPlas\altaffilmark{\ref{Washington}},
David Westman\altaffilmark{\ref{Washington}},
J.~Scott Stuart\altaffilmark{\ref{LLMIT}},
Andrew C. Becker\altaffilmark{\ref{Washington}},
Gregor Srdo\v{c}\altaffilmark{\ref{Gregor}},
Przemyslaw Wozniak\altaffilmark{\ref{LANL}},
Hakeem Oluseyi\altaffilmark{\ref{FIT}}
}

\altaffiltext{1}{Observatoire astronomique de l'Universit\'{e} de Gen\`{e}ve, 51 chemin des Maillettes, CH-1290
Sauverny, Switzerland\label{Geneve}}
\altaffiltext{2}{University of Washington, Department of Astronomy, P.O.~Box
                 351580, Seattle, WA 98195-1580, USA\label{Washington}}
\altaffiltext{3}{Department of Physics, Faculty of Science, University of
Zagreb, Bijeni\v{c}ka cesta 32, 10000 Zagreb, Croatia\label{PMF}}
\altaffiltext{4}{Hvar Observatory, Faculty of Geodesy, Ka\v{c}i\'{c}eva 26,
10000 Zagreb, Croatia\label{OH}}
\altaffiltext{5}{Faculty of Geodesy, Ka\v{c}i\'{c}eva 26, 10000 Zagreb,
Croatia\label{Geo}}
\altaffiltext{6}{Division of Physics, Mathematics and Astronomy, Caltech,
                 Pasadena, CA 91125, USA\label{Caltech}}
\altaffiltext{7}{ISDC Data Centre for Astrophysics, Universit\'{e} de Gen\`{e}ve, chemin d'Ecogia 16, CH-1290 Versoix, Switzerland\label{ISDC}}
\altaffiltext{8}{Lincoln Laboratory, Massachusetts Institute of Technology,
                 244 Wood Street, Lexington, MA 02420-9108, USA\label{LLMIT}}
\altaffiltext{9}{Sar\v{s}oni 90, 51216 Vi\v{s}kovo, Croatia\label{Gregor}}
\altaffiltext{10}{Los Alamos National Laboratory, 30 Bikini Atoll Rd., Los
Alamos, NM 87545-0001, USA\label{LANL}}
\altaffiltext{11}{Florida Institute of Technology, Melbourne, FL
32901, USA\label{FIT}}

\begin{abstract}We describe the construction of a highly reliable sample of
$\sim$7,000 optically faint periodic variable stars with light curves obtained
by the asteroid survey LINEAR across 10,000 deg$^2$ of northern sky. The majority of
these variables have not been cataloged yet. The sample flux limit is several
magnitudes fainter than for most other wide-angle surveys; the photometric
errors range from $\sim$0.03 mag at $r=15$ to $\sim$0.20 mag at $r=18$. Light
curves include on average 250 data points, collected over about a decade. Using
SDSS-based photometric recalibration of the LINEAR data for about 25 million
objects, we selected $\sim$200,000 most probable candidate variables with $r<17$
and visually confirmed and classified $\sim$7,000 periodic variables using
phased light curves. The reliability and uniformity of visual classification
across eight human classifiers was calibrated and tested using a catalog of
variable stars from the SDSS Stripe 82 region, and verified using an
unsupervised machine learning approach. The resulting sample of periodic LINEAR
variables is dominated by 3,900 RR Lyr\ae{} stars and 2,700 eclipsing binary
stars of all subtypes, and includes small fractions of relatively rare
populations such as asymptotic giant branch stars and SX~Phoenicis stars.  We
discuss the distribution of these mostly uncataloged variables in various
diagrams constructed with optical-to-infrared SDSS, 2MASS and WISE photometry,
and with LINEAR light curve features. We find that combination of light curve
features and colors enables classification schemes much more powerful than when
colors or light curves are each used separately. An interesting side result is a
robust and precise quantitative description of a strong correlation between the
light-curve period and color/spectral type for close and contact eclipsing
binary stars ($\beta$ Lyr\ae{} and W UMa): as the color-based spectral type
varies from K4 to F5, the median period increases from 5.9 hours to 8.8 hours.
These large samples of robustly classified variable stars will enable detailed
statistical studies of the Galactic structure and physics of binary and other stars,
and we make them publicly available.
\end{abstract} 

\keywords{variable stars: general --- pulsating:
RR Lyr\ae{}, $\delta$~Scuti, SX
Phoenicis, Mira, long period, semi-regular --- binaries: eclipsing ---
astronomical databases: catalogs, surveys, classification }

\section{Introduction\label{sec:intro}}

Variability is an important phenomenon in astrophysical studies of structure and
evolution, both stellar, Galactic and extragalactic. Its importance will only
increase with the advent of massive time domain surveys, such as Gaia
\citep{Ey12} and LSST \citep{ive08}, where the expected number of identified
variable stars will reach hundreds of millions -- roughly the same as the number
of all the stars detected by the Sloan Digital Sky Survey
\citep[SDSS;][]{yor00}. Such a large number of light curves can be fully
analyzed only using automated machine learning methods
\citep[e.g.,][]{deb2007,dub11, rich2011}. Most such methods require reliable training
samples; in addition to astrophysical motivation for improved understanding of
the optical variability of faint sources, a goal of analysis presented
here is to construct a large training sample of periodic variable stars that
probes both a large sky area and faint magnitude range. 

This paper is the third one in a series based on light curve data collected by
the LINEAR (Lincoln Near-Earth Asteroid Research) asteroid survey in the period
roughly from 1998 to 2009. In the first paper \citep[hereafter Paper
I,][]{ses11} we described the LINEAR survey and photometric recalibration based
on SDSS stars acting as a dense grid of standard stars. In the overlapping
$\sim$10,000 deg$^2$ of sky between LINEAR and SDSS, photometric errors range
from $\sim$0.03 mag for sources not limited by photon statistics to $\sim$0.20
mag at $r=18$ (here $r$ is the SDSS $r$ band magnitude). LINEAR data provide
time domain information for the brightest 4 magnitudes of SDSS survey, with 250
unfiltered photometric observations per object on average (rising to $\sim$500
along the Ecliptic). The public access to the recalibrated LINEAR data,
including over 5 billion photometric measurements for about 25 million objects
(about three quarters are stars; $\sim$5 million objects have $r<17$ and
photometric errors below about 0.1 mag) is provided through the SkyDOT Web site
(https://astroweb.lanl.gov/lineardb/). Positional matches to SDSS and 2MASS
\citep{skr06} catalog entries are also available for the entire sample. In this
work we also provide positional matches to WISE catalog entries
\citep{WISEpaper1} for confirmed periodic variables.

In Paper I we compared LINEAR dataset to other prominent contemporary wide-area
variability surveys in terms of depth and cadence. LINEAR extends the deepest
similar wide-area variability survey, the Northern Sky Variability Survey
\citep{woz04}, by 3 mag. This improvement in depth is significant; for example,
it can be used to extend distance limit for Galactic structure studies based on
RR Lyr\ae{} stars by  a factor of 4 \citep[to about $\sim$30 kpc; for details see the second
paper in this series, hereafter Paper II][]{ses13}. 
Thanks to the improved faint limit, the sample includes over a thousand quasars 
\citep[for $r<17$; for detailed analysis see ][]{Ruan12}. The large sky area, with resulting increase in sample sizes,
enables robust statistical studies of samples such as eclipsing binary stars,
and searches for rare objects (e.g., field SX~Phe stars, asymptotic giant branch
stars). In addition to these specific programs, the depth improvement of 3 mag
will help quantify the variation of the composition of the 
variable source population with depth. For example, \citet{EB2005} determined that
83\% of variable objects with $V<14$ are red giants, while in contrast
\citet{ses07} found that two thirds of variable objects with $14<V<20$ are RR
Lyr\ae{} and quasars). 

In order to make scientific use of the LINEAR dataset, the completeness and
purity for samples of selected variable objects need to be understood and
quantified. There are a number of automated methods for selecting variable
objects and classifying their light curves proposed in the literature
\citep[e.g.,][and references therein]{EB2005, deb2007, dub11, rich2011}.
Measuring the performance of these methods on LINEAR dataset requires reliable
training sample and full understanding of the photometric error distribution. It
would be difficult to quantify the performance of these methods on LINEAR
dataset because there are no reliable training samples, and the photometric
error distribution is not fully understood yet. The LINEAR survey was not
designed as a photometric survey, and more importantly, it accepted data
obtained in non-photometric conditions. Although the LINEAR photometric error
distribution obtained in Paper I is close to Gaussian, various tests show that
of the order 1\% of measurements can have anomalous errors (defined here as
errors at least three times larger than reported errors) that are hard to
recognize using available metadata (such as photometric zeropoint information
and the photometric scatter for calibration stars). This problem could be
explained by acquisition of data in non-photometric conditions (e.g. thin
clouds or haze). A part of the problem may also be the fact that a large fraction of
observations are obtained along the Ecliptic where contamination by blended main
belt asteroids is not negligible.

Despite the fraction of measurements with anomalous errors as small as 1\%,
the resulting sample contamination can be substantial.  According to
\citet{ses07}, about 2\% of objects with $14<V<20$ are variable at the 0.05
mag level (root-mean-square scatter, rms). Given that practical cutoff on rms is
about 0.1 mag for the LINEAR dataset, and excluding quasars which are not
numerous at magnitudes probed by LINEAR (fewer than 0.1\% of objects in the
LINEAR sample  with $r<18$ are quasars), robustly detectable variability is  
expected for much less than 1\% of the sample.  Hence, even if only 1\% of the 
LINEAR sample is spuriously selected as variable star candidates, the resulting
false positives would dominate the sample. 

LINEAR observing strategy produces repeat photometry data for stars on several
timescales, ranging from 15-20 minute interval between images within a frameset,
to a few days between repeat visits during one lunation, to the month-long
timescale between lunar months, to the yearly. More details on the sampling
pattern can be found in Paper I Appendix A.

In order to better understand the behavior of photometric errors in the LINEAR
sample, and to ultimately enable deployment of automated methods for selecting
variable objects and classifying their light curves, we have undertaken an
extensive program of visual classification of about 200,000 light curves by
eight human classifiers. Further details about visual classification and the construction
of the resulting sample of about 7,000 robust periodic variables are described in \S2. 
The distribution of periodic variables, dominated by roughly equal fractions of RR Lyr\ae{} 
stars and eclipsing binary stars, in various color-color and other diagrams is discussed 
in \S3. We compare our results to existing variable star catalogs in \S4, and to 
supervised and unsupervised machine learning classification methods in \S5. 
Our main results are discussed and summarized in \S6.

\section{Visual Classification of LINEAR Light Curves\label{sec:vis_class}}

The main goal of our analysis is the selection of a large robust sample of
periodic variable stars, with a high purity (i.e., low contamination) within
adopted flux, amplitude and period limits. To improve the sample robustness and
light curve classification, we undertook three successive selection and
classification steps. After the initial sample selection, period estimation and
construction of phased light curves, eight human classifiers extracted about
7,000 likely periodic variables from a starting set of about 200,000 candidate
variables, and also obtained initial light curve classification. In the
following two steps, a single expert refined selection and classification of the
smaller sample of 7,000 likely periodic variables, first by repeating visual
classification, and then further refined the candidate sample by adding the
parameters measured from light curves and other information, such as multicolor
photometry into the classification procedure. In this section we first describe
the initial sample selection and period estimation, and then discuss the visual
classification procedures in detail. A preliminary analysis of the resulting
sample of robust periodic variables is presented in the next section. 

\subsection{Sample selection\label{sec:selection}}

We start by selecting candidate variables from the public LINEAR 
database\footnote{Available at https://astroweb.lanl.gov/lineardb/}
using the following criteria: 

\begin{itemize}
\item Brightness limit: $14.5 < \langle m_{LINEAR} \rangle < 17$, where $\langle m_{LINEAR} \rangle$
is the median value of the white-light LINEAR magnitude\footnote{The faint magnitude limit
adopted in Paper II is 0.5 mag fainter than adopted here because ab~type~RR~Lyr\ae{} discussed
in Paper II are easier to recognize than other types of variable object discussed here.}.
\item Likely variability:  $\chi^2_{dof} > 3$, where $\chi^2$ per degree of freedom is 
computed using the unweighted mean magnitude and photometric errors reported in 
the database. 
\item Variability amplitude: $\sigma > 0.1$ mag, where $\sigma$ is the rms
scatter (standard deviation) of recalibrated LINEAR magnitudes. 
\end{itemize} 

The majority of about 200,000 selected objects are found in the region bounded
by $125^\circ < {\rm R.A.} < 268^\circ$ and $-13^\circ < {\rm Dec} < 69^\circ$
(corresponding to the North Galactic Cap scanned by SDSS). Additional
$\sim$8,000 objects are found in the SDSS Stripe 82 region ($-50^\circ < {\rm
R.A.} < 60^\circ$ and $|{\rm Dec}| < 1.266^\circ$). The selected objects contain
both true variable objects and spurious candidates. We limit our classification
to objects exhibiting mono-periodic variability (light curves $m(t)$ that satisfy
$m(t+P) = m(t)$, where $P$ is the period and $t$ is positive; assuming no
noise), and use phased light curves for visual inspection. Phased light curves
are constructed by plotting $m(t)$ as function of phase
\begin{equation}
              \phi =   {t \over P} - {\rm int}\left( {t \over P} \right),
\end{equation}
where the function int($x$) returns the integer part of $x$. The likely periods were 
determined as described next.

\subsection{Period finding methods}\label{period finding}

For each selected object, the  three most likely periods were found using using
an implementation of the Supersmoother algorithm \citep{fri84, rie94}. This
non-parametric method smooths the light curve using a variable smoothing length
and uses cross-validation method to pick a best-fit period with the smallest
phased light curve dispersion. The Supersmoother algorithm was extensively used
by the MACHO survey and should be robust for a large variety of variable stars
because it makes no explicit assumptions about the shape of the light curve.

During the classification it soon became apparent that the Supersmoother algorithm
often had problems with finding the correct period; for eclipsing binaries in particular 
a large fraction of best-fit periods were twice as short as the true period (we will
return to this discussion in \S\ref{compareP}). For this reason, we also included two
additional algorithms for estimating periods: the Lomb-Scargle (LS) and Generalized 
Lomb-Scargle (GLS) parametric methods \citep{lomb76,scargle82,zk09}. We used the
code implemented in Gaia's Coordination Unit 7
pipeline \citep{Ey13}. 

The LS method essentially  fits a single sine wave to the light curve, and is
capable of using heteroscedastic errors. It assumes that the true light curve
mean is equal to the mean of sampled data points. In practice, the data often do
not sample all the phases equally, the dataset may be small, or it may not
extend over the whole duration of a cycle: the resulting error in the estimated
light curve mean can cause problems such as aliasing. A simple remedy
implemented in the GLS algorithm is to add a constant offset term to the single
sinusoid model \citep{zk09}. 

We note that when the light curve shape significantly differs from a single
sinusoid, the LS and GLS methods may easily fail. Possible remedies in such
cases are to fit pre-defined light curve templates \citep[e.g.,][]{ses10}, or to
use multiple harmonics in the Fourier expansion, which we have not considered
here (e.g. Figures \ref{fig:period_finding} and \ref{fig:LSproblem}). 

\subsection{Visual classification methodology}
\label{subsection:vis_class_method}

Visual classification was performed on a per-object basis. There were three 
classification/validation runs; the first run pruned the list of candidates by more than
a factor of 20, and the subsequent two runs further improved the sample 
purity and light curve classification precision. In the first run, 200,000 variable 
star candidates were divided roughly equally among eight human classifiers, using 
right ascension boundaries, and each classifier processed approximately 30,000 light 
curves. Overlaps of 2,500 light curves between the samples of the ``adjacent'' classifiers 
were used to verify classification consistency (which was assessed as described in 
\S\ref{classtests} and \ref{S82tests}). 

\subsubsection{Initial visual classification}

\begin{figure}
\epsscale{1}
\plotone{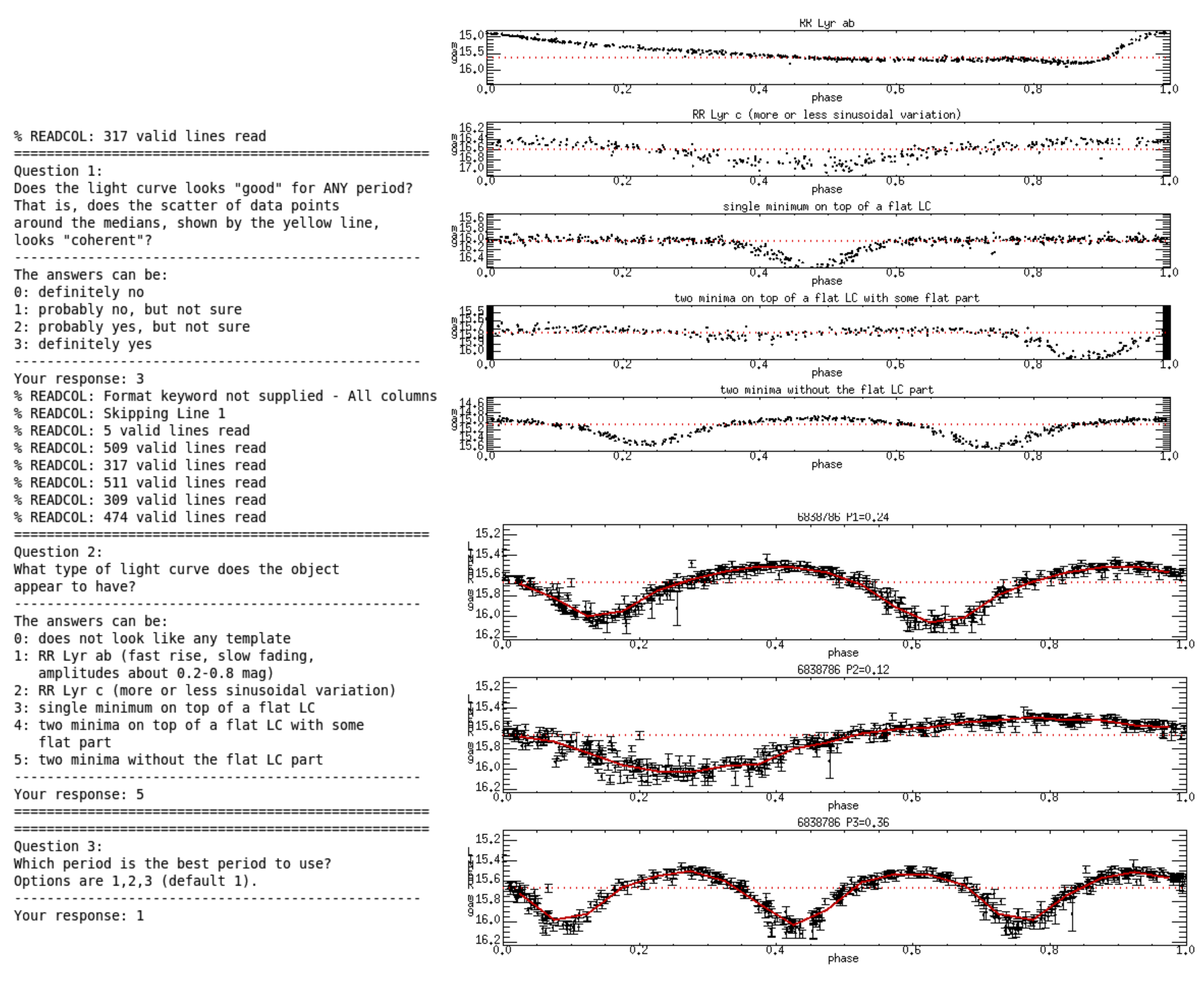}
\caption{ User interface for the classification tool. The three bottom right
panels show phased LINEAR light curves of the given object for the three most
probable periods calculated by the Supersmoother algorithm. The five top right
panels represent light curve templates used in classification.
\label{fig:VisCode}
}
\end{figure}

The initial visual classification was performed using the user interface
shown in Figure~\ref{fig:VisCode}.  The automated classification tool displayed three phased 
light curves, folded with the periods found by the Supersmoother period finding algorithm, as 
well as five templates of folded (phased) light curves spanning predicted classes of variable
objects. Classifiers answered three questions with fixed possible answers. 

The first question was whether the displayed phased light curves have ``reasonably small''
dispersion around some imaginary smooth shape, following the Phase Dispersion
Minimization idea of \citet{ste78}. There were four possible answers to this
question (coded by numerical values in parentheses): ``definitely no'' (0),
``probably no, but not sure'' (1), ``probably yes, but not sure'' (2), 
``definitely yes'' (3). Unless the answer to the first question is ``definitely
no'', classifier proceeds to the second question related to the light curve
shape. Possible answers are: ``does not look like any template'' (0), ``RR Lyr
ab'' (1), ``RR Lyr c'' (2), ``single minimum on top of a flat light curve'' (3),
``two minima on top of a flat light curve with some flat part'' (4), ``two
minima without the flat light curve part'' (5). The third question asks the user
to choose which of the three folded light curves of the given object shows the
smallest dispersion (the intention was to determine which of the three periods
is the best). In addition, there was an option to add comments if necessary
(e.g., about period aliasing, or any problems with the data), or to go back and
repeat the classification for the object if an error was made. By design, {\it only
the light curve shape was used in this first classification stage.} 

After a brief training period, it takes about 5 seconds on average to answer
all three questions, for a throughput of $\sim$700 objects per hour (about a
week worth of full-time work per classifier, or about 2 Full-Time-Equivalent
person months for the  whole effort, assumming an unrealistic efficiency of 100\%).

\subsubsection{Tests of the initial classification uniformity and repeatability \label{classtests}}

In order to assess the uniformity and repeatability of the visual classification, 
a subsample of 8,044 light curves was classified by all eight classifiers. These 
objects were selected from the SDSS Stripe 82 region so that a comparison with 
an SDSS-based variable object catalog can also be performed (described further below). 

For each light curve, we averaged the eight answers to question 1 (ranging from 0 for 
``definitely not variable'' to 3 for ``definitely variable'') to obtain its $A1$ ``grade''. 
We also computed its standard deviation among the eight classifiers, $\sigma_{A1}$, 
to quantify dispersion in classification grades. Based on the morphology of the $A1$ 
distribution, we divided the sample into four subsamples using $A1$, as summarized in
Table~\ref{tab:tableA1}. The 317 light curves with $A1 > 1.8$ have the smallest $\sigma_{A1}=0.15$:  
that is, most classifiers agree that these 3.9\% objects are ``definitely variable''. 
The classification robustness of other light curves is lower, as seen from the 
increased dispersion among the classifiers. 

After sorting light curves by $A1$, two coauthors have re-inspected all 438
light curves with $A1 > 1.1$ (classes 1-3), as well as 1000 light curves from
class 0 with highest $A1$ values. No spurious classifications were found in class 3. 
Objects in class 2 seem definitely variable, but many appear to have incorrect
periods. Class 1 is similar to class 2, except for a larger fraction of unconvicing 
periodic cases. Therefore, there are between 317 and 438 definite periodic
variables in this sample, depending on how conservative a selection cut is adopted,
implying an upper limit for the sample contamination of 28\%. Our main
conclusion is that human classifiers are mutually consistent when their answer
to the first classification question is 2 or 3, that is, when they are highly confident 
about detected variability.

\subsubsection{Robust $\chi^2$ Selection} 
 
The LINEAR light curve database contains two values of $\chi^2$: the standard
value and the so-called robust $\chi^2$, $R\chi^2$, determined by excluding both
brightest and faintest 10\% of points from the computation (note that despite
its name, the measured $\chi^2$ does not follow the statistical $\chi^2$
distribution expected for Gaussian photometric errors). The robust $\chi^2$
might be efficient at minimizing the impact of photometric outliers, but at the
same time it may decrease the sample completeness for light curves where
variability is not always present (e.g., bursts and Algol-like light curves). 
 
We have investigated whether $R\chi^2$ can be used to significantly prune the
initial sample without a large decrease in the final sample completenesss (that is,
whether  $R\chi^2$-based selection could be used instead of visual pruning of
the candidate sample). If $R\chi^2 > 3$ selection is adopted (instead of $\chi^2 > 3$), the size of the
initial sample decreases from $\sim$200,000 to $\sim$80,000. Of all the light
curves with $A1 > 1.2$ (classes 2 and 3 above, see Table~\ref{tab:tableA1}), 86\% 
have $R\chi^2 > 3$. Therefore, the initial sample could be made smaller by a factor 
of 2.5, while losing 10-20\% of true variables. This tradeoff reflects both the 
properties of faint variable stars and the behavior of LINEAR photometry. 
 
About 14\% of light curves with $A1>1.2$ (robust variables, as suggested by
visual classification) have $R\chi^2 < 3$ (no strong evidence for variability).
We have re-inspected these puzzling cases and found that they all are indeed
real variables. In other words, visual classification is correct but $R\chi^2 < 3$ 
is too conservative a cut -- these objects mostly have small amplitudes,
short-duration peaks, or are faint (and thus photometric errors are large).
Therefore, it should be possible to extract additional variable stars from the
LINEAR database because our initial sample of 200,000 candidates had to
satisfy $\chi^2>3$. 
 
We have also re-inspected a random sample of light curves with $A1 < 1.2$ and
$R\chi^2 > 3$, that is, light curves that show significant variability according to 
$R\chi^2$ but were not visually classified as periodic variables. About a half of these
light curves show significant variability which appears aperiodic. A subset of a few
hundred light curves with periods exceeding 1000 days and $R\chi^2 > 10$ seem
consistent with being semi-regular variable asymptotic giant branch stars. 
Therefore, their rejection from the periodic light curve sample during visual
classification is justified. 

In summary, $R\chi^2$ parameter cannot be used to replace the visual classification
step by automated selection without a significant drop in the sample completeness.

\subsubsection{Comparison to the variable star sample from the SDSS Stripe 82\label{S82tests}}

SDSS has obtained multiple observations (about 50 on average) in the $300 deg^2$ large
so-called Stripe 82 region. These data were used to select 67,507 candidate variable point
sources\footnote{Light curves are publicly available from \\
http://www.astro.washington.edu/users/ivezic/sdss/catalogs/S82variables.html}
(for details see \citealt{ive07a}; \citealt{ses07}; and references therein).
There are many more candidate variables per unit sky area in the SDSS Stripe 82
catalog than in LINEAR sample because the former is much deeper ($g<20.5$ vs.
$r<17.5$) and has a more inclusive cutoff for variability rms (0.05 vs. 0.1 mag). We
have used this SDSS catalog to assess the reliability and completeness of
candidate variables visually selected from LINEAR database. 
 
\begin{figure}
\epsscale{0.5}
\plotone{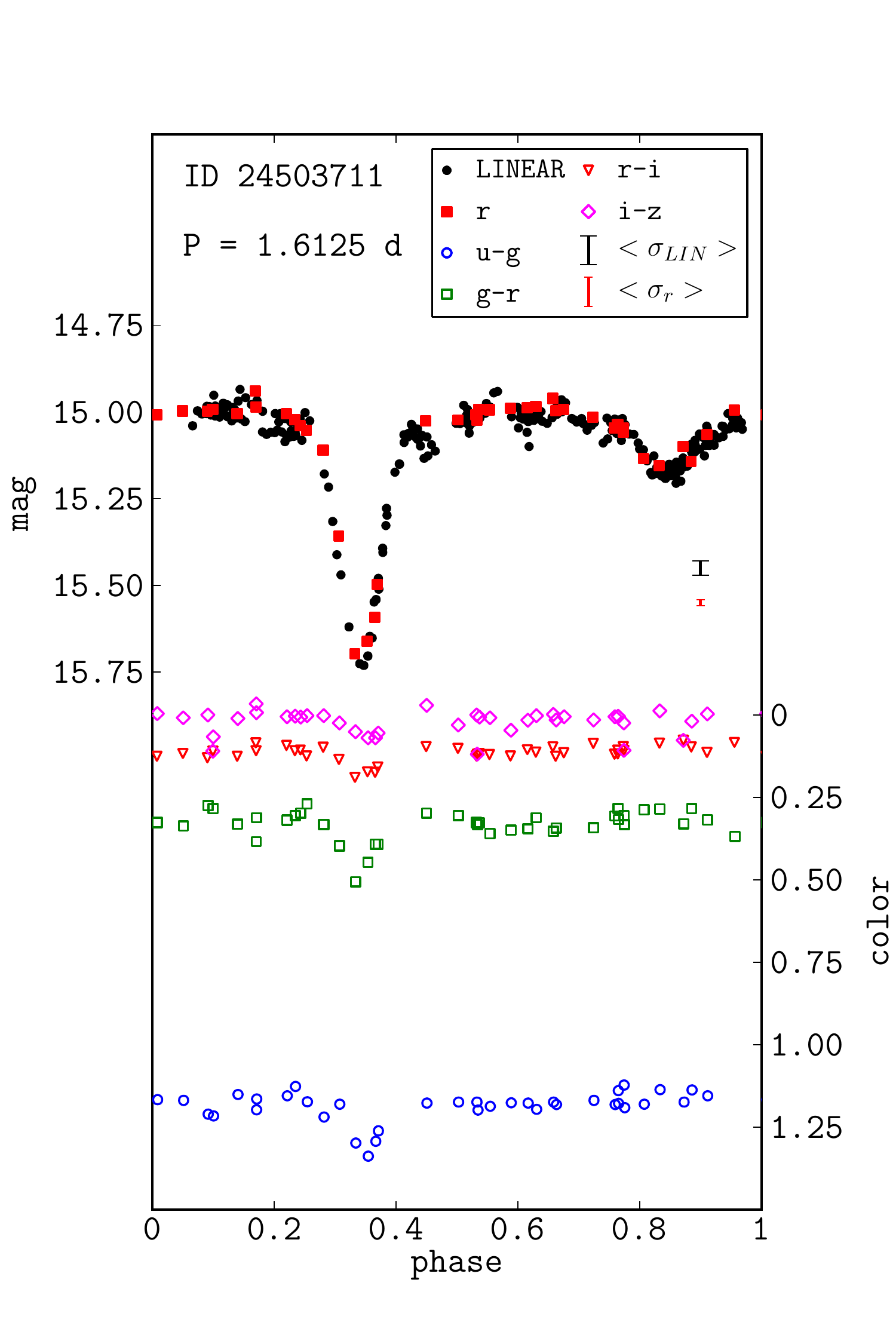}
\caption{Example of LINEAR/SDSS synergy. Scale on the left corresponds to
unfiltered LINEAR magnitudes and scale on the right to SDSS colors. ``Top and
bottom (black and red in the online version) bars show average LINEAR and SDSS
errors, respectively''. LINEAR provides a better cadence for studying variable
objects, while SDSS provides multi-band photometry that encodes valuable
additional information about the variable object. \label{fig:nice_algol}
}
\end{figure}

Out of 8,044 LINEAR objects found the Stripe 82 region, 543 have positional matches 
within 2 arcsec to candidate SDSS variables that show periodic behavior. Of those,
301 have $A1>1.2$, that is, 83\% of 363 robust LINEAR variables are confirmed by
SDSS data. Therefore, there are 62 robust LINEAR variables that are not in SDSS variable
sample, representing an 11\% addition to the SDSS sample. These 62 LINEAR variables 
are dominated by detached eclipsing binaries with most SDSS observations falling along 
the flat part of light curve. An example is shown in Figure~\ref{fig:nice_algol}. Therefore, 
the implied purity of $A1>1.2$ LINEAR variables must be higher than 83\%, and is consistent
with 100\% (that is, we did not find a single questionable case among these 62 variables). 
Figure~\ref{fig:nice_algol} also demonstrates synergy between the SDSS and LINEAR datasets: 
while LINEAR provides much better time-resolved photometry for studying variable objects, 
SDSS provides very informative 5-band photometry. 

About 45\% of SDSS variables which are sufficiently bright to be in LINEAR sample
are not selected from LINEAR database using criteria listed in \S\ref{sec:selection} and 
$A1>1.2$ based on visual classification. About one third of those could be recovered by 
relaxing the $A1$ limit. The remaining two thirds ($\sim$30\% of all SDSS variables) 
typically have sparse LINEAR data and/or small variability amplitudes, and thus were
justifiably rejected in visual classification. Therefore, relative to the SDSS subsample
limited to a similar depth, the completeness of the LINEAR sample is in the range 55-70\%, 
depending on the adopted $A1$ cut (most of the LINEAR incompleteness is due to larger 
adopted minimum rms variability, 0.1 mag vs. 0.05 mag). 

Finally, out of 301 stars that are recognized as periodic variables by both SDSS and
LINEAR, 184 have LINEAR and SDSS periods that agree within 2\%. Additional 57
objects have periods aliased by a factor of 2 in either SDSS or LINEAR  (for one
third of those, the SDSS periods are larger); they include a large fraction of eclipsing
binary systems with similar depths of primary and secondary minima. 

\subsubsection{Iterative improvements to visual classification}

The first classification step, which pruned the initial list of  200,000 candidate variables by 
more than a factor of 20, was performed by eight different classifiers which must have 
introduced some non-uniformity in the resulting classification. In addition, the resulting
sample contamination could be as high as 17\%, as discussed in \S\ref{classtests} and 
\S\ref{S82tests}. To improve sample purity and classification 
uniformity,  all the objects tagged as plausibly variable in the first round 
were re-examined in the second round by the first author. Only a few percent of
objects had their classification changed as a result of this re-examination. Generally,
no significant variations among the eight subsamples were noticed, in agreement
with the conclusions from the previous sections. 

When the available source attributes (period, amplitude, and skewness of light curves,
and optical and infrared colors) were analyzed for the sample obtained in the second 
classification round, it became apparent that different types of variable stars cluster
in different regions of the multi-dimensional attribute space. Using selection boundaries 
based on color, period, amplitude and light curve skewness listed in Table~\ref{tab:boundaries},
and discussed in more detail in the next subsection (\S\ref{sec:boundaries}), 
an additional sample of about 750 objects was selected from the initial candidate sample 
of 200,000 objects. That is, about 10\% more potential variables than extracted in the first 
classification round were selected for further inspection. 

Visual inspection of these 750 candidates (by the first author) in the third
classification round revealed that only about 10\% represented convincing cases
of periodic variability. They were added to the initial list to produce the
final sample of 7,194 visually selected and classified periodic variables. Among
those, 6,876 light curves (96\%) have been assigned a definite type, while the
remainder are classified as ``Other''. The latter group contains objects which
are variable, but not periodically and objects for which the exact variability
type could not be reliably determined. 

The six main light curve types are listed in Table~\ref{tab:boundaries}, and a
few supplemental ones in Table~\ref{tab:results_VC}, and discussed in more
detail in the next Section. Hereafter, we refer to this entire sample as
``visually confirmed sample of periodic LINEAR variables'', or simply ``PLV''
sample. The resulting catalog is made publicly available\footnote{Available from
http://www.astro.washington.edu/users/ivezic/r\_datadepot.html}. 

Table \ref{tab:results_VC} quantitatively summarizes the results of visual
classification. The first column ``translates'' our numerical codes used during
visual classification to the adopted variability types. We hypothesize that the
class ``3'' (``a single minimum on top of a flat light curve) mostly consists of
EA type binaries (Algols) for which our data did not show a discernible
secondary minimum (i.e. either too shallow to be detected, or too similar in
depth to the primary minimum, recall \S2.3). For that class of objects correct
periods could be twice longer than listed in the catalog. The  light curves
classified as ``5'' include two types of eclipsing binaries: EB (or $\beta$
Lyr\ae{}) and EW (W Ursae Majoris), which are grouped together because they are
hard to distinguish using only LINEAR light curves. Motivated by the
distribution in period-color and period-amplitude diagrams, we introduced two
additional classes: class ``6'' (containing SX~Phoenicis and $\delta$~Scuti
candidates) and class ``7'' (long-period variables defined here as variables
with periods longer than 50 days, and as semi-regular variables). Further
explanations regarding introduction of these two additional classes can be found
in \S\ref{sec:SXPhe} and \S\ref{sec:wise}.

\subsubsection{Simple automated classification with the aid of other attributes \label{sec:boundaries}}

The clustering of objects in different regions of the multi-dimensional attribute space
offers an opportunity to develop automated classification methods. Here we define 
selection boundaries using simple, rectangular cuts in the four-dimensional
attribute space (period, amplitude, skewness, $g-i$ color). Alternative approaches
based on machine learning algorithms are discussed in Section \S\ref{sec:automClass}. 
The adopted boundaries are listed in Table~\ref{tab:boundaries}. We limit quantitative analysis of the 
performance of this classification scheme to ab and c~type~RR~Lyr\ae{}, EB/EW eclipsing 
binaries and SX~Phoenicis/$\delta$~Scuti candidates. We do not include classes whose size
does not exceed 1\% of the full sample, nor Algols (EA eclipsing binaries) and objects
classified as ``Other''. We do not include Algols because their distribution does not
have well-defined boundaries (not too surprising since in the case of detached 
binaries we could easily have an ensemble of paired objects with presumably few common physical 
characteristics). An analogous diversity is expected among long-period variables
which include both Miras and semi-regular variables, and possibly other classes
of variable stars. Indeed, even the definition of Mira stars suffers from quantitative
ambiguity (``red long-period variables with visual amplitudes exceeding 2.5 mag''), 
although it has been shown that they are actually fundamental mode pulsators --- 
a physical characteristic that differentiates them from other long period variables
\citep[e.g.][]{woo96,sos09,span11}. 

\begin{figure}
\vskip -0.3in
\epsscale{0.6}
\plotone{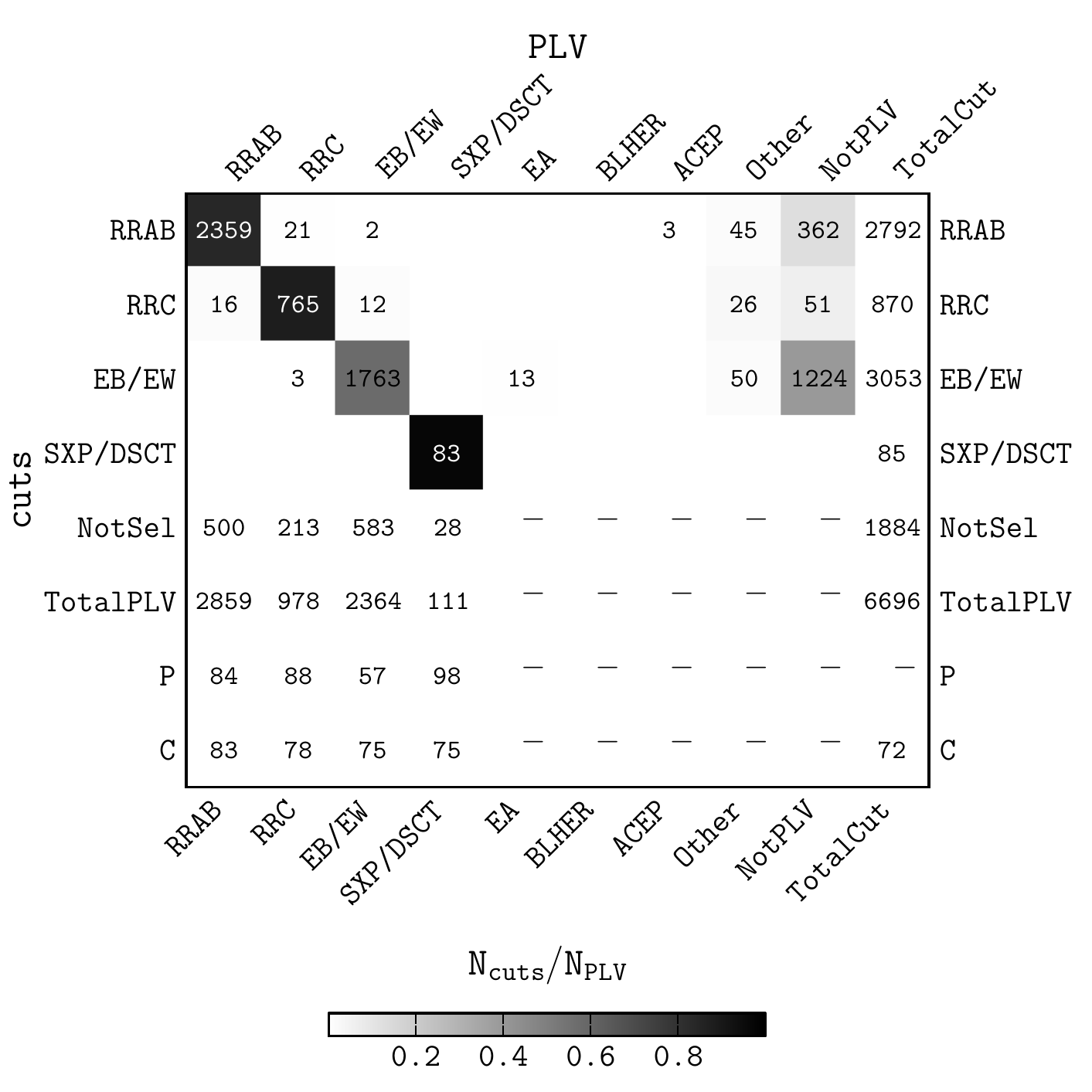}
\vskip -0.1in
\caption{Statistical performance comparison between the visually confirmed and
classified variable sample listed in the final PLV catalog (7,194 objects), and
a simple supervised classification algorithm aplied to the full sample of all
200,000 candidate variables. The selection boundaries for the latter are listed
in Table \ref{tab:boundaries}. The columns correspond to light curve types used
in the PLV catalog; in addition, the column labeled ``Other'' corresponds to
variable PLV objects that do not belong to any of other chosen variability
types, and the ``NotPLV'' column corresponds to objects that satisify the
selection cuts applied to the full sample but that were not visually tagged as
variable and included in PLV. The first four rows correspond to the four
analyzed subsamples of variables defined by applied selection cuts. The last
column lists the total number of objects selected by each automated cut. The
fifth row, labeled ``NotSel'', corresponds to PLV objects not selected by
automated selection cuts, and the sixth row, labeled ``TotalPLV'', gives the sum
of the fifth row and the number of PLV objects of a given type correctly
classified by automated method. The intersection regions are color-coded by the
fraction of objects in each row falling into a given region, that is the
fraction of selected objects with type confirmed by PLV, that are also listed in
the penultimate row. The last row, labeled ``C'', lists the completenes of the
automated selection method when compared to PLV for the four analyzed
variability classes. \label{fig:att-cuts-class} 
}
\end{figure}

In order to maintain analysis uniformity, we use best-fit periods found by the
classic Lomb-Scargle method. Objects with unreliably measured SDSS colors, and
Lomb-Scargle periods close to one day and half a day ($\pm$0.05 tolerance in
log$P$) were excluded from the analysis. The performance of this supervised
classification is statistically compared to our  visual classification results
in Figure~\ref{fig:att-cuts-class}. We have visually re-examined all 3,270 light
curves with differing visual and automated classifications. 

The automated method selected 74\% of PLV objects from the four analyzed types.
This result does not imply a 26\% contamination in the PLV catalog but rather 
an incompleteness of the automated selection method; the majority of missing 
objects had unreliable SDSS colors, were rejected by the period cut, or had at least
one of the attributes outside the allowed interval. This selection fraction varies little 
among the four types (see the bottom row in Figure~\ref{fig:att-cuts-class}). 

The automated selection method selected additional 835 objects that are not included
in the PLV catalog (a 12\% addition, varying from 4\% for c~type~RR~Lyr\ae{} to
23\% for EB/EW). Of those 835 objects, 246 correspond to ab~type~RR~Lyr\ae{};
the majority are located very close to the red cutoff 
for the $g-i$ color.  Approximately 15\% of these 246 objects have light curves 
hinting at ab~type~RR~Lyr\ae{}, but not of sufficient quality to enable reliable visual 
confirmation. Therefore, at most about 40 ab~type~RR~Lyr\ae{} included in the initial
sample of 200,000 candidates are missing from the PLV catalog (1.4\% effect).
In case of c~type~RR~Lyr\ae{}, 44 objects not 
in PLV are uniformly distributed throughout the selection volume. About
30\% of these objects have light curves that might be classified as c type
RR Lyr\ae{}, though not reliably. Similar behavior is displayed in EB/EW case,
with only about 10\% of 545 objects not in PLV potentially classifiable
as reliably periodic. Therefore, the PLV catalog is only slightly incomplete
relative to the initial sample of 200,000 candidates (by about 1-2\% at most). 

The automated classification is correct for a high fraction of PLV objects:
97\% for ab~type~RR~Lyr\ae{}, 78\% for c~type~RR~Lyr\ae{},  87\% for EB/EW, 
and 100\% for SX~Phe/$\delta$ Sct. In summary, this analysis provides further support that 
the PLV catalog is highly complete relative to the initial sample of 200,000 
candidate variables, has exceedingly low contamination, and a high rate of 
correct light curve classification.

\subsubsection{Comparison of period finding methods \label{compareP}} 

As we already indicated earlier, period finding algorithms often had problems with choosing
the correct period. For example, for eclipsing binaries a large fraction of best-fit 
periods were twice as short as the true period. In this particular case, such behavior
is easy to understand: primary and secondary minima are often of similar depth and are 
therefore often misidentified as the same feature in the phased light curve. This error, 
however, is not seen consistently: not all of the objects with similar depths of minima have 
periods that are too short by a factor of two.

\begin{figure}
\epsscale{0.5}
\plotone{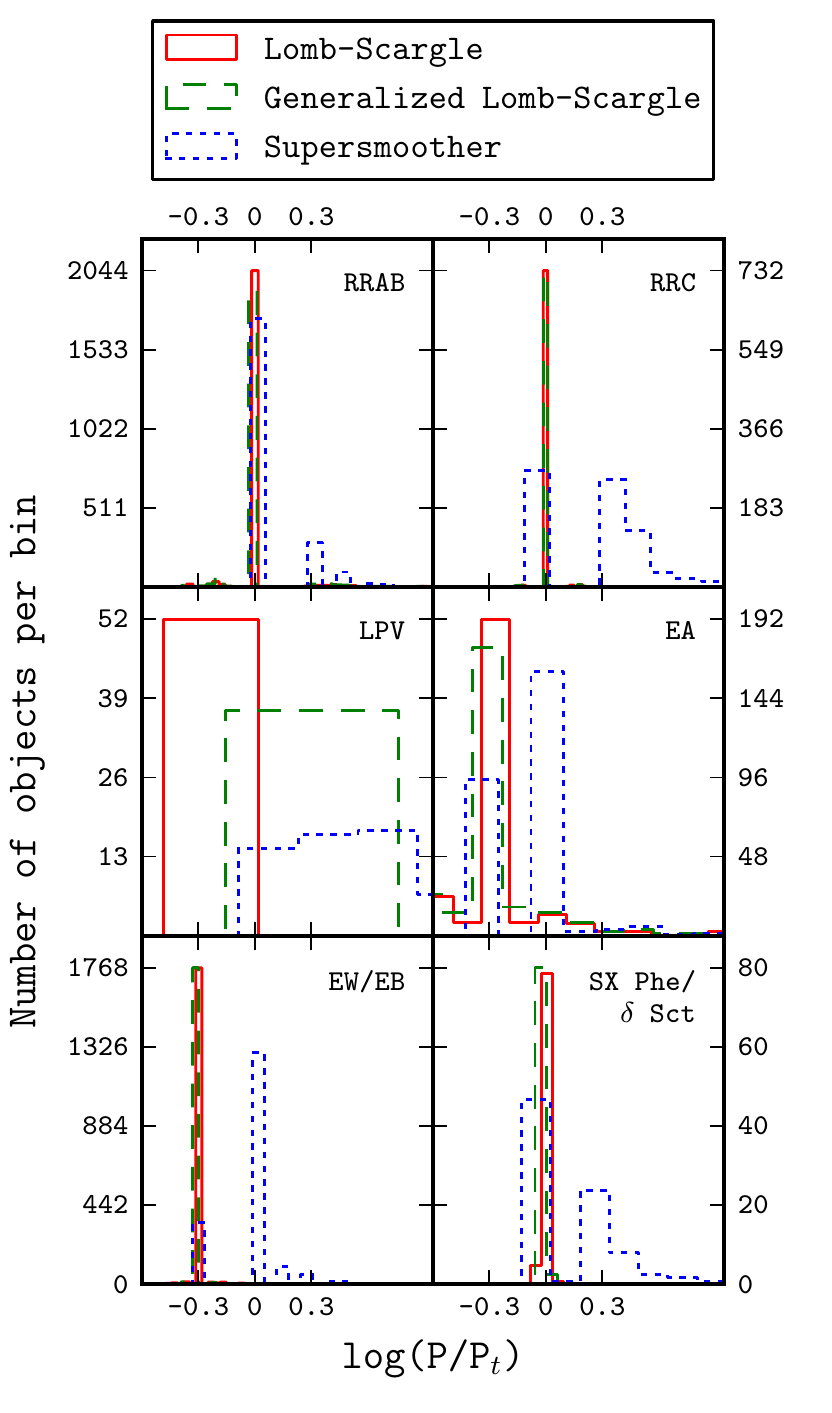}
\caption{ Comparison of the three period finding methods, separately for each of
the six main light curve types (clockwise, from the top left:
ab~type~RR~Lyr\ae{}, c~type~RR~Lyr\ae{}, EA eclipsing binaries (Algol),
SX~Phe/$\delta$~Sct variables, EW/EB eclipsing binaries ($\beta$ Lyr and W UMa),
and long-period variables (asymptotic giant branch stars). The abscissa shows
the logarithm of the ratio of the period computed by each method and the
visually confirmed true period (note that a factor of 2 bias corresponds to 0.30
on logarithmic scale). Note that the Lomb-Scargle methods consistently
underestimate period of EA and EW/EB light curves by a factor of 2 (this
systematic effect has been corrected in the public catalog). Color version of
this figure is available online. \label{fig:period_finding}
}
\end{figure}

\begin{figure}
\epsscale{1.0}
\vskip -2.5in 
\plotone{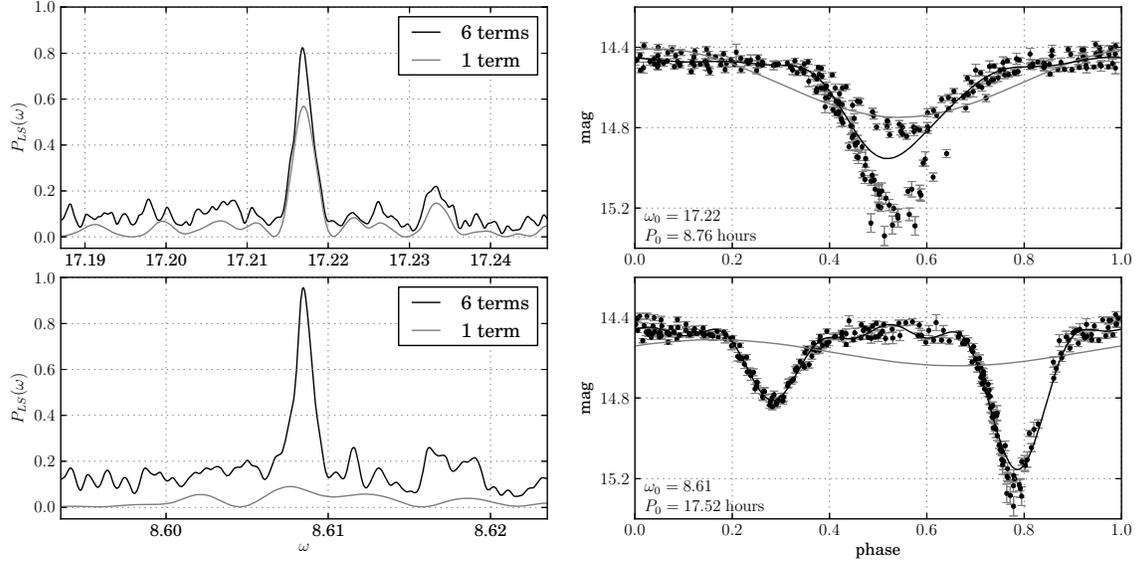}
\vskip -2.5in 
\caption{An illustration of the failure of the Lomb-Scargle method to find the
correct period when the light curve shape significantly differs from a single
sinusoid. The two top panels show the Lomb-Scargle periodogram (left) and phased
light curves (right) for truncated Fourier series models with one and six terms.
Symbols with error bars represent LINEAR data for star with ID=14752041 (the
data and the python code to produce this figure, including period estimation,
are publicly available from the {\it astroML} site, http://astroml.github.com).
Phased light curves are computed using the aliased period favored by the
single-term model, and the model light curves are shown by lines using the same
line styles as in the top left panel. The correct period is favored by the
six-term model but unrecognized by the single-term model, as illustrated in the
bottom left panel. The phased light curve constructed with the correct period is
shown in the bottom right panel. This figure is adapted from \citet{ive13}, and
can be reproduced using code available at http://www.astroML.org \citep{van12}.
\label{fig:LSproblem}
}
\end{figure}

Given the final sample of 6,876 reliably classified light curves,  we tested period 
finding  methods for each of the six main light curve types separately. Our results 
are summarized in Figure~\ref{fig:period_finding}. We left the ``single
minima on top of a flat light curve'' class out of the analysis, as the sample
is small (20 objects) and the correct period for those objects could not be
identified with certainty. We speculate that those objects could correspond to
eclipsing binaries of EA (Algol) type with similar depths of minima, but with
periods that are too short by a factor of two. Another explanation would be that
secondary minima for these objects are too shallow to be detected in LINEAR data. 

Our results show that the Lomb-Scargle and generalized Lomb-Scargle methods 
typically outperform the Supersmoother algorithm for all variability types. For c type 
RR Lyr\ae{}, long-period variables, and SX~Phe/$\delta$~Sct type light curves, Supersmoother 
has a much larger fraction of overestimated periods (typically by a factor of two, but 
sometimes more) than the other two methods. In addition, when the period is 
approximately correct, the uncertainty is typically larger for Supersmoother values
(that is, the width of the central peak in histograms shown in Figure~\ref{fig:period_finding}
is larger). 

The performance of the period finding algorithms for eclipsing binaries is
rather different: while the Lomb-Scargle and generalized Lomb-Scargle methods
produce narrower histogram peaks than Supersmoother, their periods are 
consistently (at $>$90\% level) too short by a factor of two! After an overall
correction of periods for eclipsing binaries by this factor, the  Lomb-Scargle 
and generalized Lomb-Scargle methods display better performance than Supersmoother. 

The reason for this consistent bias in period estimation by the Lomb-Scargle and generalized 
Lomb-Scargle methods is their fundamental assumption that the shape of the underlying
light curve can be described by a single sinusoid. A remedy is to fit a Fourier series with many 
terms (but more computationally expensive). As illustrated in Figure~\ref{fig:LSproblem},  a 
Fourier series model with six terms correctly recognizes two minima in the light curve of an 
eclipsing binary star. For additional discussion, please see \cite{hoffman09} and \cite{wyr03}. 

During the visual inspection it was relatively easy, albeit time consuming, to apply this correction 
factor to the periods. In a fully automated classification scheme that has only single band light 
curves and no color information this might be more difficult since values of period, amplitude 
and skewness are in large part similar for c~type~RR~Lyr\ae{} and EB and EW binaries. Addition of 
appropriate color information (e.g. $g-i$) easily breaks this degeneracy (see \S3.1 and \S3.2).
Ultimately, the performance of period finding algorithms based on a single sinusoid
can be significantly improved by including more Fourier terms.

\section{Analysis of Periodic LINEAR Variables \label{sec:analysis}}

The remainder of our analysis is performed using the public version of the PLV catalog. 
We show in this section that the distribution of selected periodic variables displays
distinctive features in the multi-dimensional attribute space spanned by the light-curve
parameters (period, amplitude, shape) and optical/infrared colors. This behavior enables 
robust and efficient classification of objects into various classes of variable population.
These features are not  seen for the full sample of 200,000 candidate variable objects, 
and thus strongly suggest that visual classification successfully extracted true variables. 

We first discuss the distribution of classified variables in diagrams constructed with 
the three light curve parameters, and then investigate the correlation of light curve 
parameters with optical and infrared colors. We quantify a strong correlation between
the period and optical color for contact eclipsing binaries, provide evidence that 
the sample contains a large number, compared to the known objects, of likely 
Population II field SX~Phe stars, and demonstrate that the infrared colors from the
WISE survey provide further support that long-period variables are correctly classified.

\subsection{Analysis of Light Curve Properties}

The light-curve amplitude is estimated non-parametrically from the cumulative
magnitude distribution as the range between the 5\% and 95\% points. The
light-curve skewness is computed as described in \citet{ses07}. Therefore, light
curves are quantitatively described using three parameters: period, amplitude
and skewness. This choice is of course not unique. For example, in addition to,
or instead of, amplitude, other estimators of the width of the observed
magnitude distribution could be used, such as standard deviation (which is not
robust to outliers) and the inter-quartile range (which, depending on the
sampling, might not be sensitive to single minima in otherwise flat light
curves). Similarly, the light-curve shape could be further quantified using
higher moments (such as kurtosis, but they quickly become very noisy), Fourier
coefficients (which help greatly to classify eclipsing binary subtypes
\citep{poj02}, or RR Lyrae subtypes \citep{sos11}), or even
non-parametrically using the principal component analysis
\citep[e.g.][]{ds2009}. In this preliminary analysis, we find that even our
simple approach based on period, amplitude and skewness provides informative
description of the light curve behavior. Nevertheless, exploring these other
options would be a worthwhile analysis to undertake. 

\begin{figure}
\epsscale{1}
\vskip -0.2in
\plotone{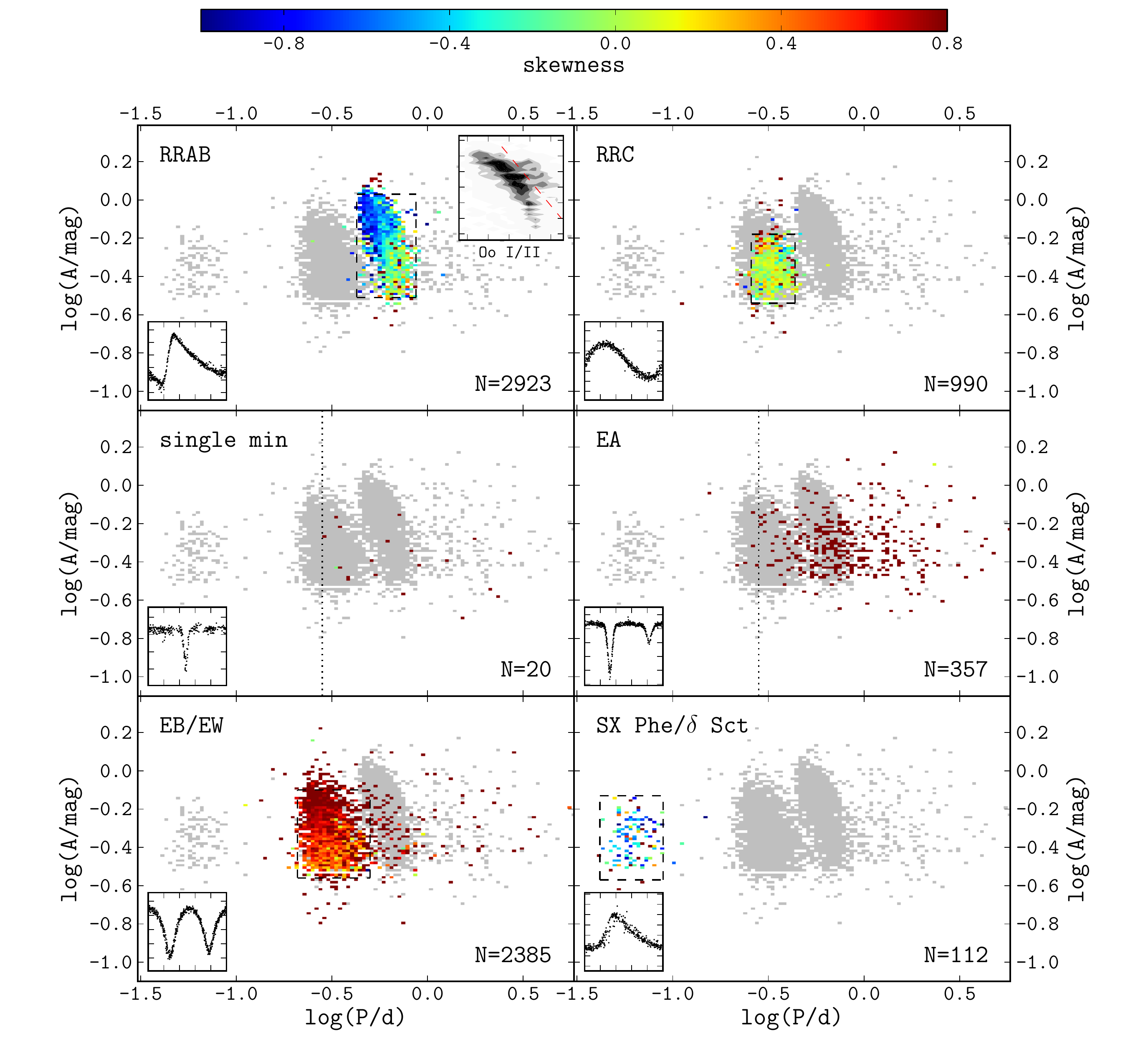}
\vskip -0.2in
\caption{Period---Amplitude diagram for visually confirmed periodic LINEAR
variables (PLV). Each panel represents a given class of variable stars confirmed
by visual classification. Width of the bins is 0.03 in color and 0.02 in
$log(P/d)$. Bins are color coded by the median value of skewness (per bin). Grey
backgroud corresponds to all PLV sample variables. Insets in panels represent a
typical light curve for the variability type in the given panel. N is the total
number of objects of a given type. ``Top right inset in the ``RRAB'' panel shows
separation between Oosterhoff type I and type II RR Lyr\ae{} ab, former being
left and below dashed line (red dashed line in the online version) and the
latter right and above". The gray map shows the density of ab~type~RR~Lyr\ae{}
per bin. Axes in the folded light curve diagrams correspond to phase and
magnitude, and in the inset in the Oosterhoff diagram for RR Lyr\ae{} to
$log(P/d)$ and $log(A/mag)$. Color version of this figure is available online.
\label{fig:logP-logA-skew}
}
\end{figure}

\begin{figure}
\epsscale{1}
\plotone{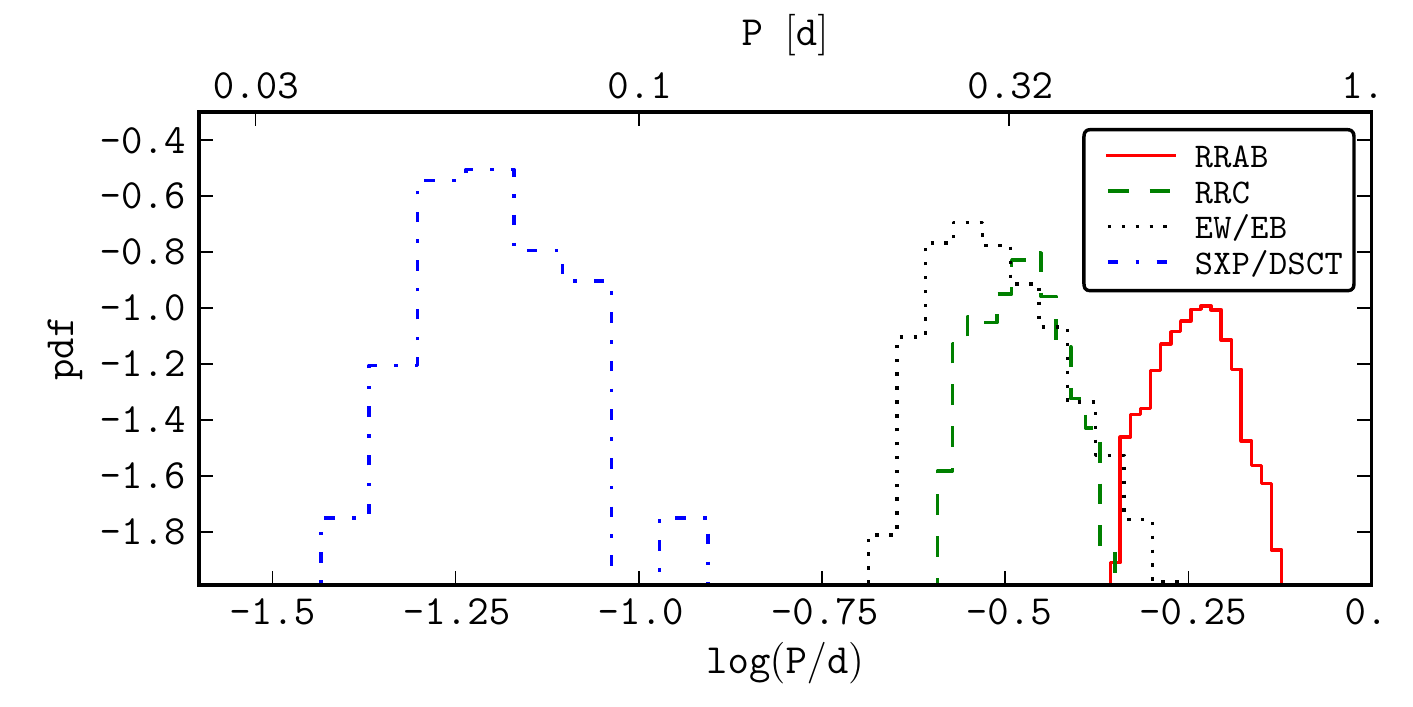}
\caption{The period distribution for the five most populous variability classes.
While SX~Phe/$\delta$~Sct candidates clearly stand out ($P<0.1$ day), and ab
type and c~type~RR~Lyr\ae{} are fairly well separated by $P=$0.4 days, eclipsing
binaries overlap with the period range of RR Lyr\ae{} stars (especially EW/EB
type eclipsing binaries and c~type~RR~Lyr\ae{}). Color version of this figure is
available online. \label{fig:per_class_histo}
}
\end{figure}

The distribution of variables in the period--amplitude--skewness space 
is illustrated separately for each of the six main variability classes in 
Figure~\ref{fig:logP-logA-skew}. The period distribution of the PLV sample is 
multi-modal, as further quantified in Figure~\ref{fig:per_class_histo}. Even the 
period alone enables remarkable, although not perfect, classification of periodic 
variables: SX~Phe/$\delta$~Sct candidates clearly stand out ($P<0.1$ day), and 
ab type and c~type~RR~Lyr\ae{} are fairly well separated by $P=$ 0.4 days. Nevertheless, 
eclipsing binaries overlap with the period range of RR Lyr\ae{} stars (especially EW/EB 
type eclipsing binaries and c~type~RR~Lyr\ae{}). In addition, the light-curve amplitude
distributions are similar for c~type~RR~Lyr\ae{} and EB/EW eclipsing binaries. 
This degeneracy can be readily lifted using the light curve skewness (and object color, 
see below). Indeed, all six classes can be readily defined when all three
light-curve parameters are considered (e.g. EB/EW class has much larger 
skewness than c~type~RR~Lyr\ae{}; compare the symbol color in the top right
and bottom left panels in Figure~\ref{fig:logP-logA-skew}). In other words,
the visual classification of light curves in essence reflects the distribution 
of these three parameters (and also of the light curve smoothness). We analyze the
performance of automated classification methods based on this behavior in 
\S\ref{sec:automClass}. 

It is possible to further separate ab~type~RR~Lyr\ae{} into Oosterhoff type I and
Oosterhoff type II stars \citep{ses10}, as shown in the top right inset in the
``RRAB'' panel of Figure~\ref{fig:logP-logA-skew} (note also the strong correlation
between the amplitude, skewness and period for ab RR Lyr\ae{}). Average periods of 
Oosterhoff type I and type II ab RR Lyr\ae{} for the PLV sample are $\langle P_{ab}^I \rangle =
0.56$ days and $ \langle P_{ab}^{II} \rangle = 0.65$ days. This result is in good
agreement with Oosterhoff's conclusion that period of RR Lyr\ae{} ab in Oosterhoff
type I clusters is 0.1 day shorter than that of those in Oosterhoff type II clusters
\citep{OO44}.  For a more detailed analysis of the Oosterhoff's dichotomy for 
field RR Lyr\ae{} stars based on this sample, see \cite{ses13}.

\subsection{Correlations between Colors and Light Curve Properties}

The addition of the color information to light-curve parameters significantly improves
the separation of visually defined classes and ultimately enables better performance 
of automated classification methods. For a detailed discussion of the distribution of 
stars in various color-color diagrams constructed with SDSS and 2MASS photometry, 
see \citet{cov07}, and references therein. The most useful SDSS-2MASS colors are 
$u-g$, $g-r$ (or $g-i$), $i-K$ and $J-K$, which are sensitive to various combinations 
of effective temperature, metallicity, and surface gravity. Therefore, the minimal useful 
dimensionality (the number of measured attributes that are independent for at least
some subsamples) of this dataset is at least five (the three light curve attributes and at 
least two color attributes). 

We emphasize that both SDSS and 2MASS photometry are
single-epoch measurements obtained at random light curve phases.  Therefore, 
while the observed color range tracks the intrinsic color range of a given population,
distribution of objects {\it within} that range is affected by the color light curve shape
(e.g. ab~type~RR~Lyr\ae{} stars spend more time close to minimum than to maximum light; 
since RR Lyr\ae{} are redder when fainter, their instantaneous color distribution is skewed
redwards compared to their mean color distribution). 

\begin{figure}
\epsscale{1}
\plotone{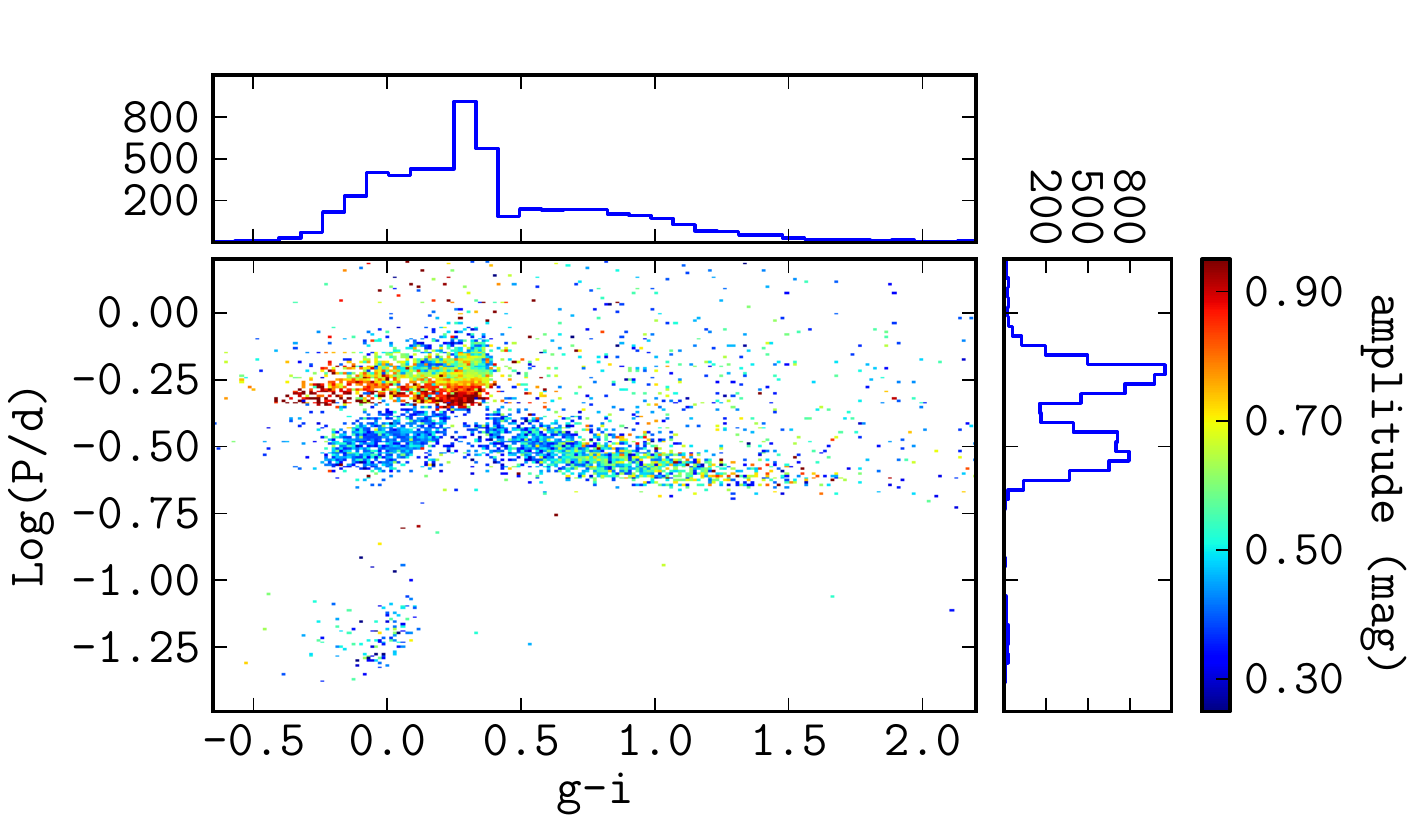}
\caption{The distribution of periodic variables in the period-color diagram.
Bins are color coded by the median value of light curve amplitude according to
the legend on the right. The two histograms show marginal distributions of the
period and the $g-i$ color. Color version of this figure is available online.
\label{fig:gi-logP-A_pc_hist}
}
\end{figure}

\begin{figure}
\epsscale{1}
\vskip -0.2in
\plotone{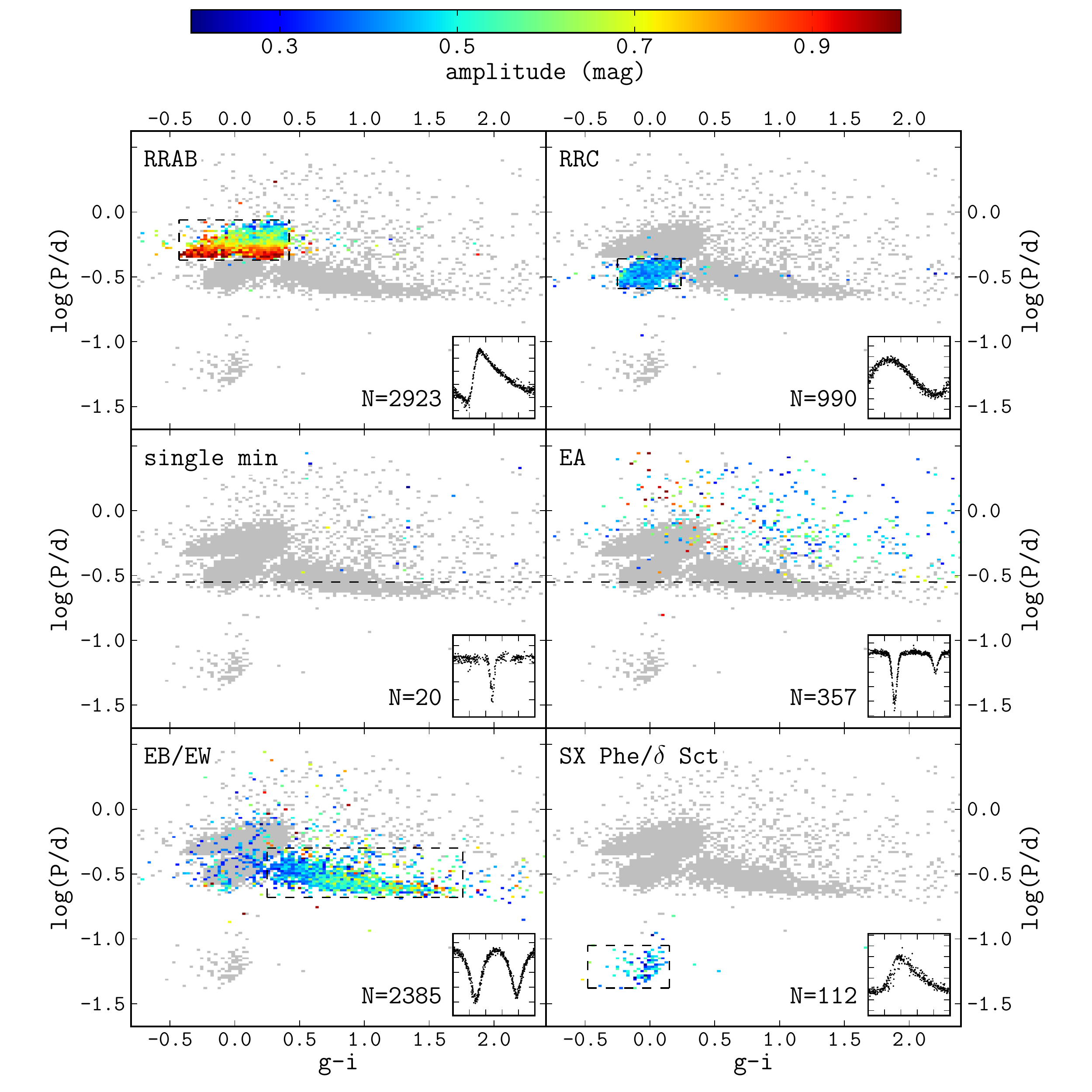}
\vskip -0.2in
\caption{The distribution of visually confirmed periodic LINEAR variables (PLV)
in the period-color diagram.  Each bin has been color coded by the median
amplitude of objects inside it, according to the color bar above. Width of the
bins is 0.03 in color and 0.02 in $log(P/d)$. Grey backgroud represents all PLV
sample variables. Insets in panels represent a typical light curve for the
variability type in the given panel. N is the number of objects of a given type.
Axes in the folded light curve diagrams correspond to phase and magnitude. The
dashed lines outline the selection boundaries listed in
Table~\ref{tab:boundaries} and discussed in \S\ref{sec:boundaries}. Color
version of this figure is available online. \label{fig:gi-logP-A_pc} 
}
\end{figure}

Figure~\ref{fig:gi-logP-A_pc_hist} demonstrates that the addition of just one color to 
the period, here the SDSS $g-i$ color which is a good measure of the effective temperature 
\citep{tomoII}, helps to clearly separate c~type~RR~Lyr\ae{} from EB/EW binaries. A more
detailed illustration of the correlations between the $g-i$ color and light curve properties
is shown in Figure~\ref{fig:gi-logP-A_pc}. Note in particular how EA and EB/EW are well
separated in this diagram. The EB/EW subsample displays a good correlation between
the period and color, discussed in more detail in \S\ref{PC_binaries}.

\subsubsection{The $g-r$ vs. $u-g$ diagram}

In addition to the three-dimensional $g-i$ color--period--amplitude projection
of the full multi-dimensional attribute space discussed above, the
three-dimensional projection spanned by the SDSS $u-g$ and $g-r$ colors and
light curve skewness is also rich in content. The $u-g$ vs. $g-r$ diagram is one
of the most informative SDSS color-color diagrams; it clearly distinguishes
quasars from stars, main sequence stars from binary stars and white dwarfs, and
it contains information about effective temperature and even metallicity for
blue main sequence stars \citep{smo04,ive07a,tomoII}. 

\begin{figure}
\epsscale{0.9}
\vskip -0.5in
\plotone{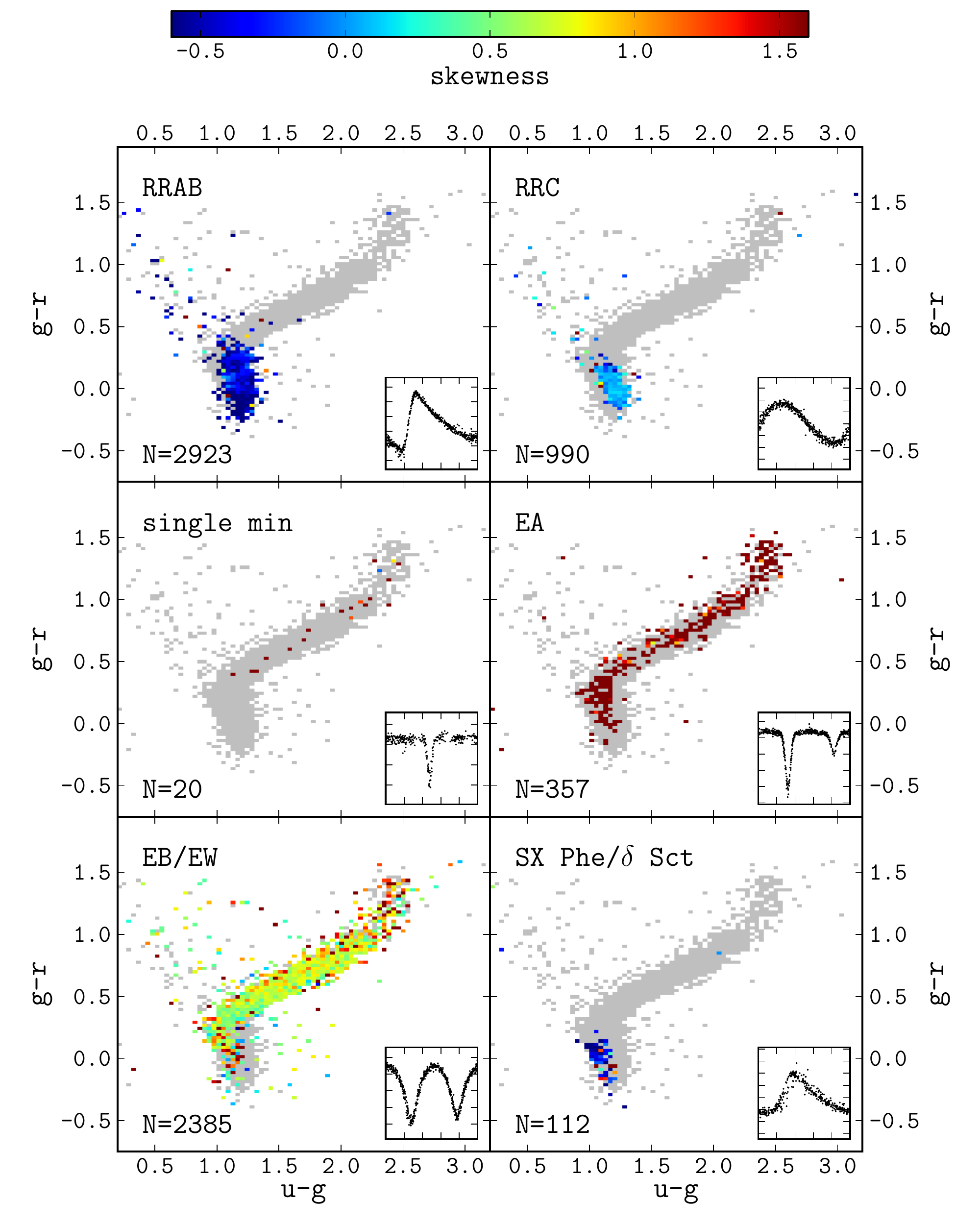}
\vskip -0.1in
\caption{Analogous to Figure~\ref{fig:gi-logP-A_pc}, except that the PLV sample
distribution is shown in the SDSS $g-r$ vs. $u-g$ color-color diagram, and the
color coding is based on the light curve skewness. Color version of this figure
is available online. \label{fig:ug-gr-skew} }
\end{figure}

The distribution of variables in the $u-g$ vs. $g-r$ vs. skewness space is shown 
separately for each of the six main variability classes in Figure~\ref{fig:ug-gr-skew}. 
As known from previous work based on SDSS data, RR Lyr\ae{} color distribution
is localized to the region populated by spectral types A and early F \citep[][and 
references therein]{ses10}. Only about 1-2\% of light curves classified as 
RR Lyr\ae{} fall outside the expected small color regions discernible in Figure~
\ref{fig:ug-gr-skew}. 

Based on the $u-g$ vs. $g-r$ color-color diagram and the skewness distributions,
we identified approximately 25\% suspected misclassifications between c type RR
Lyr\ae{} and EB/EW eclipsing binaries (from the first classification round) and
visually re-inspected their light curves. We found that approximately 80\% of
those were indeed likely misclassifications and their type was subsequently
revised. The cross-contamination of these two
subsamples is easy to understand; a light curve of an eclipsing binary with
similar depths of minima can easily be misidentified as a nearly symmetric
(sinusoidal) c~type~RR~Lyr\ae{} light curve. This ambiguity is particularly
problematic in case of faint objects, or objects with sparsely sampled light
curves. We note that the color distribution of c type  RR Lyr\ae{} has a well
defined red edge -- it is thus easy to prevent the contamination of EB/EW
subsample by c~type~RR~Lyr\ae{} but the converse is not true because EB/EW stars
can have colors as blue as RR Lyr\ae{} colors. 

We have also explored a few other three-dimensional projections of the
seven-dimensional attribute space (there are 35 possible independent attribute
combinations) and did not find diagrams as revealing as the $g-i$ color vs.
period vs. amplitude diagram and the $g-r$ vs. $u-g$ vs. skewness diagram. 
A noteworthy color is the 2MASS $J-K$ color which is capable of separating 
main sequence stars from quasars and late-type giants (including the long-period
asymptotic giant branch stars); for main sequence stars the 2MASS $J-K$ color 
and the SDSS $g-i$ color are highly correlated (both are by and large driven by the
effective temperature), while for those other populations the measured $J-K$ color 
is redder than the $J-K$ color of main sequence stars of the same $g-i$ color
\citep[for more details, see][]{cov07}.

\subsection{Period-color correlation for contact eclipsing binaries\label{PC_binaries}}

\begin{figure}
\epsscale{1.0}
\vskip -1.5in
\plotone{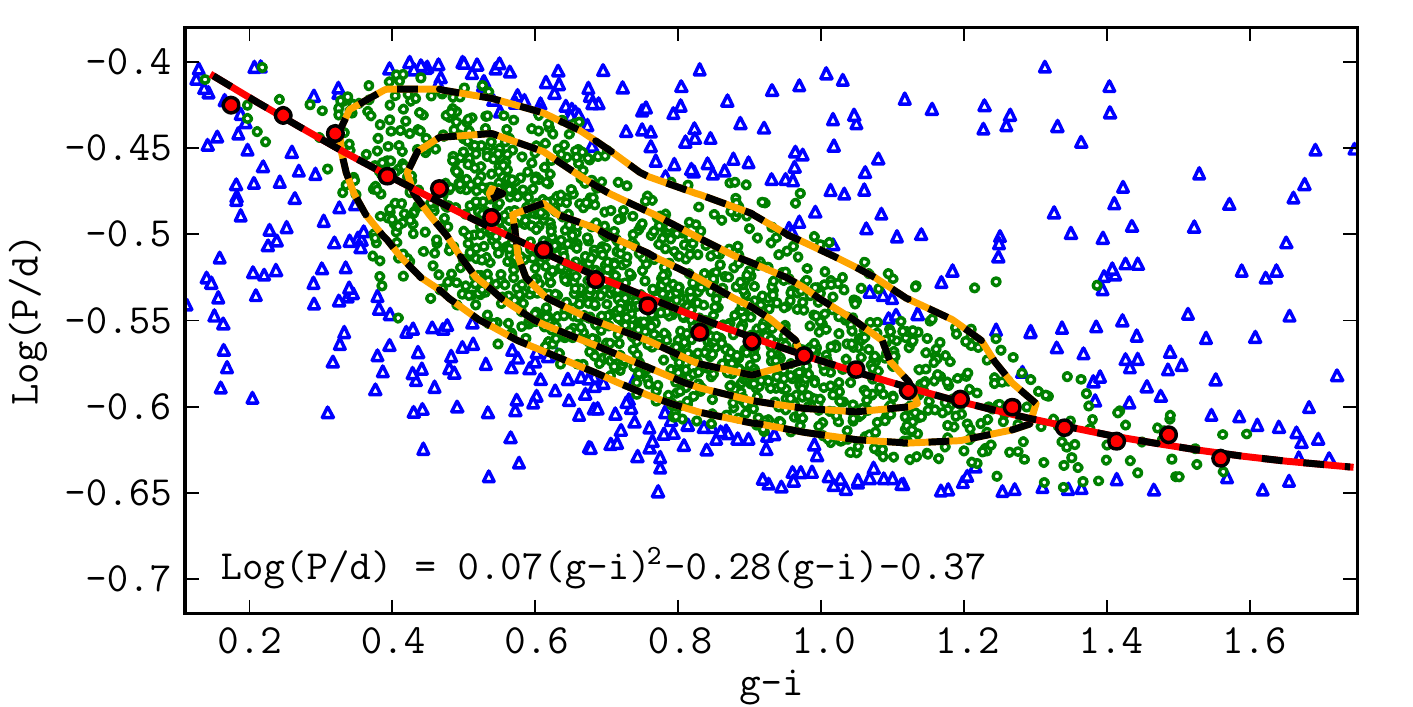}
\vskip -0.1in
\caption{Quadratic fit for correlation between period and color of EB/EW
binaries. Selected objects were visually classified as EB/EW binaries and
satisfy criteria: $0.1 < g-i < 1.8$ and  $ -0.67 < log(P) < -0.4$. Data were
binned in 0.1 wide bins in ($g-i$) and 0.05 wide bins in $log(P)$. ``Objects
outside 5\% and 95\% points of the distribution in color and period per bin were
removed from subsequent fitting procedure (triangles, blue in the online
version)". Remaining objects (circles, green in the online version) were again
binned in 0.1 wide bins in ($g-i$), and median of logarithm of period in days
per bin was calculated (filled points, red in the online version). Quadratic
function was fit to those points. Dashed contours designate areas of 5, 10, 15
points per bin, i.e. the density of objects that passed the 5\% and 95\% cuts.
Color version of this figure is available online.
\label{fig:ECfit} 
}
\end{figure}

The distribution of EB ($\beta$ Lyr\ae{}) and EW (W Ursae Majoris) eclipsing binary stars
is remarkably well outlined in the period vs. $g-i$ color diagram (see the bottom left 
panel in Figure~\ref{fig:gi-logP-A_pc}, and a zoomed-in version in Figure~\ref{fig:ECfit}).
Since the sample selection is primarily driven by the light-curve shapes, and substantial 
selection effects in the $g-i$ color and period in the relevant ranges are not expected,
this strong correlation is likely of astrophysical origin. A similar result was
reported for a much smaller sample of contact binary systems by \cite{eggen67} 
\citep[see also][and references therein]{ruc97}. The range of observed $g-i$ colors correspond 
to spectral types from F5 ($g-i=0.3$) to K4 ($g-i=1.4$) \citep[see Table 3 in][]{cov07}. 
\cite{ruc97} used Hipparcos distance estimates for 40 W UMa stars to derive a relationship 
between the absolute $V$ band magnitude, period and color. According to their results, 
our sample includes stars with $1 < M_V < 6$. 

We compute the median $log(P)$ in bins of the $g-i$ color for stars with 
$0.2 < g-i < 1.6$ and  $-0.4 < log(P) < -0.67$, and fit a parabola to the
resulting points, 
\begin{equation} 
         log(P/day)= 0.05\, (g-i)^2 - 0.24\, (g-i) - 0.37. 
\end{equation} 
Due to the large sample size, the random errors for the fitted data points are 
sufficiently small to rule out a linear relationship. This best-fit relation implies 
that the median period for EB/EW eclipsing binaries increases from 5.9 hours to 
8.8 hours as the color-based spectral type varies from K4 to F5.  An alternative
form based on the Johnson $B-V$ color, derived using using transformations 
between the SDSS and Johnson systems from \cite{ive07b}, is 
\begin{equation} 
         log(P/day)= 0.038\, (B-V)^2 - 0.29\, (B-V) - 0.33,
\end{equation} 
and valid in the range $0.3 < B-V < 1.1$. This relation agrees well with a similar
relation obtained by \cite{ruc97s} for $\sim$400 W UMa stars observed by the
OGLE project in Baade's window (note that we fit the median relation and Rucinski 
obtained the short-period limit as a function of color; the two sequences are offset
by about 0.1-0.15 mag at a given period). 

These findings are related to the fact that the period distribution for contact 
binary star systems appears to have a well-defined lower limit at 0.22 days \citep{ruc92}.
More recent data show that this limit may be a bit smaller ($\sim$0.20 days, see
\citealt{dk2010};  \citealt{dav13}), but the existance of a well-defined boundary 
is not disputed. Indeed, the falloff of the distribution at small periods for M dwarf
systems (see Figure~6 in \citealt{bec11}) is very similar to the falloff for EB/EW systems
in our Figure~6. If we extrapolate our best-fit to $g-i=2.0$ corresponding to the
spectral type M0, we obtain a period of 0.22 days in good agreement with other
studies. 

\begin{figure}
\epsscale{0.7}
\vskip -1.5in
\plotone{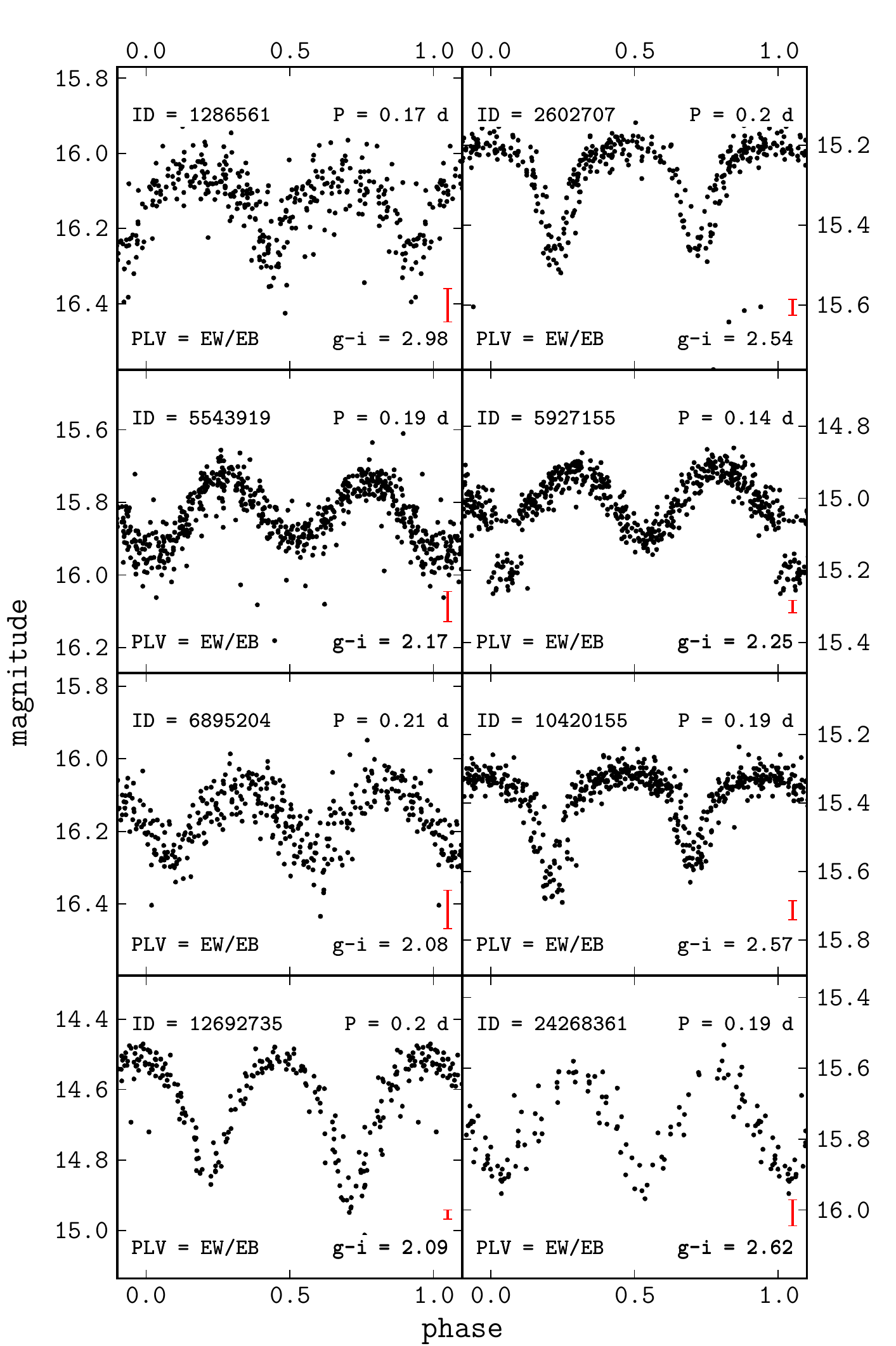}
\vskip -0.1in
\caption{Examples of short period contact binaries. Some periods are shorter
than 0.2 days and test the value of the boundary mentioned in
\S~\ref{PC_binaries}. Our most likely candidate for the eclipsing binary with
the shortest period is in the top left corner. Vertical errorbars show typical
photometric errors for each light curve. Note the unusual light curve of the
object with LINEAR ID 5927155 (follow-up in progress). \label{fig:Mdwarfs} 
}
\end{figure}

In Figure~\ref{fig:Mdwarfs} we show several examples of these short period
binaries. Several objects have periods below 0.2 days and test the value of the
aforementioned period boundary.

\subsection{Candidate SX~Phe stars \label{sec:SXPhe}} 
 
The PLV sample presented here includes a class of 112 blue stars 
($-0.3 < g-i < 0.2$, bluer than thick disk and halo turn-off stars and 
corresponding to $-0.2< B-V < 0.3$ using transformations between the 
SDSS and Johnson systems from \citealt{ive07b}), with very short periods
(1--2.5 hours), and with asymmetric light curves (see bottom right panel in 
Figures~\ref{fig:logP-logA-skew} and \ref{fig:gi-logP-A_pc}). These stars 
can be identified as a mixture of $\delta$~Scuti and SX~Phoenicis stars 
\citep[e.g. see Figure~8 in][]{EM2007}. Both types of stars are usually 
considered as variable counterparts of blue straggler stars (main sequence stars in 
open or globular clusters that appear younger than they should be given the cluster 
age), with $\delta$~Sct subsample belonging to Population I disk stars and SX~Phe 
subsample to Population II halo stars \citep[see e.g.][]{jeon2004}. 

In a recent study based on the largest catalog of SX~Phe stars assembled to 
date (about 250 stars identified in globular clusters), \cite{CohSara12}
demonstrate that this population appears to occupy a narrow region at the 
bottom of the instability strip with $1.5 < M_V < 3.5$, and are all likely radial 
mode pulsators. Given the apparent magnitude limits of our sample, 
the implied distances span the range 2--10 kpc, that is, many disk scale heights
away, and thus SX~Phe probably dominate because they are Population II (halo) objects. 
We note that the $B-V$ color distribution of our sample extends to bluer colors than the
range displayed by the \cite{CohSara12} sample (their range is approximately 
$0.1< B-V < 0.4$, corresponding to $0.0 < g-i < 0.3$; about 20\% of our candidates 
have $g-i<-0.1$). 

\begin{figure}
\epsscale{1.1}
\vskip -4.5in
\hskip -0.7in
\plotone{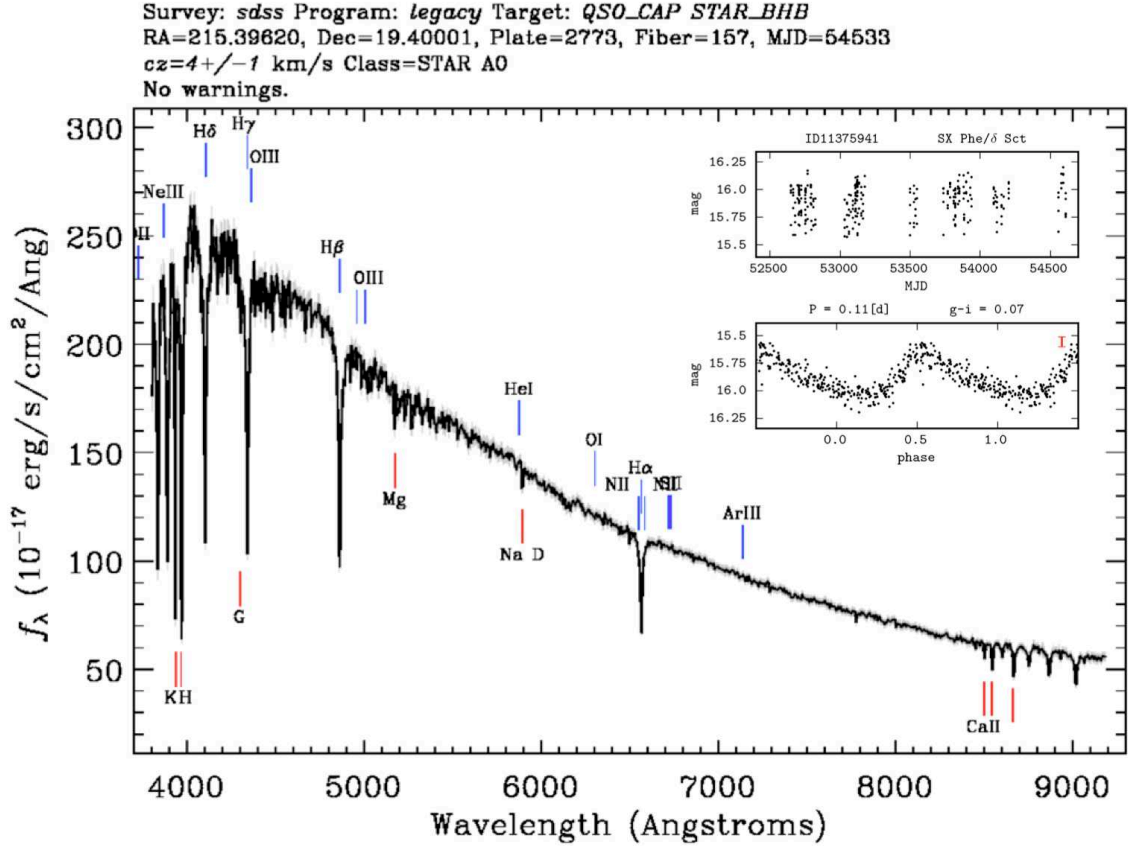}
\vskip -2.5in
\caption{Default SDSS visualization of the SDSS spectrum for an SX~Phe candidate
(LINEAR ID=11375941). The inset shows the observed and phased LINEAR light
curves. Color version of this figure is available online.
\label{fig:specSXPhe}
}
\end{figure}

\begin{figure}
\epsscale{0.9}
\vskip -0.5in
\plotone{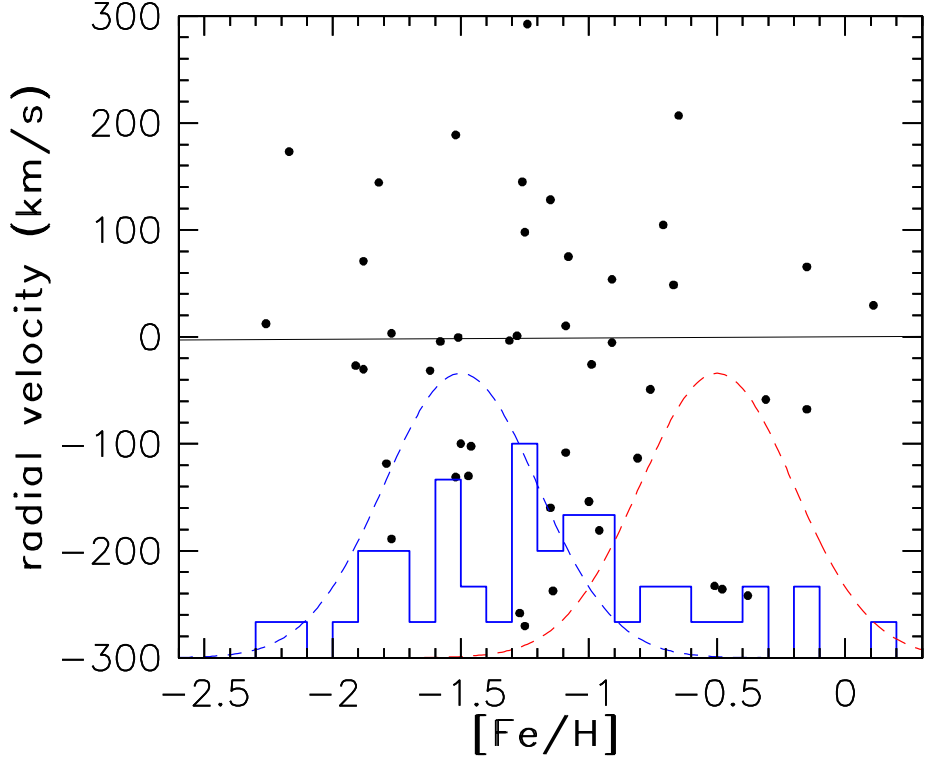}
\vskip -3.0in
\caption{The symbols show metallicity vs. radial velocity distribution of 34
candidate SX~Phe stars with SDSS spectra (repeated measurements for 7 stars are
also shown). The histogram shows the marginal distribution of metallicity. The
two Gaussians illustrate expected metallicity distributions for halo stars
(left) and disk stars (right), taken from \cite{tomoII}. The metallicity is
below the traditional boundary for separating halo and disk stars at $[Fe/H] =
-1.0$ dex for 57\% of measurements. For these stars, the radial velocity
dispersion is 135 km/s, and fully consistent with the halo hypothesis
\citep{tomoIII}. Only the four stars with $[Fe/H] > -0.5$ dex and small radial
velocity are consistent with the disk hypothesis. Color version of this figure
is available online. \label{fig:radvelFeH}
}
\end{figure}

A much higher fraction of SX~Phe stars than $\delta$~Sct stars in this sample
is supported by SDSS spectra that are available for 34 stars in the candidate sample. 
All the spectra appear very uniform and characteristic for A stars; an example is shown in Figure~\ref{fig:specSXPhe}.
Using the default SDSS metallicity and radial velocity estimates (see Figure~\ref{fig:radvelFeH}), 
we find that the sample is dominated by stars with 
[Fe/H]$<-1$, low metallicities characteristic of halo stars, with
a large velocity dispersion (134 km/s) that is also consistent with presumed halo 
population \citep[for a review of recent observational constraints on the 
differences between the metallicity and kinematics distributions of disk and 
halo stars, see e.g.][]{ibj12}. 

Assuming that our conclusion about the sample being dominated by halo 
stars is correct, these 112 candidates likely represent a major addition to the 
total number of known SX~Phe stars \citep[according to][fewer than 300 SX~Phe stars 
are known]{CohSara12}. Our sample would also increase the number of known 
{\it field} SX~Phe stars by as much as a factor of six \citep[according to][there are 
only 17 known field SX~Phoenicis known]{rod01}. This large increase in the 
sample size of field SX~Phe stars is due to the fact that the LINEAR dataset
is among the first ones to explore sufficiently faint flux levels, over a large
sky area, and with appropriate cadence. We are currently undertaking photometric
and spectroscopic followup efforts to better characterize this sample.

\subsection{Candidate AGB stars and WISE color distribution \label{sec:wise}}

The PLV sample includes 77 light curves classified as ``long-period 
variables'', defined here as variables with periods longer than 50 days, and 
as semi-regular variables. These stars are expected to be dominated by 
asymptotic giant branch (AGB) stars which often display infrared excess
emission due to their dusty envelopes \citep[see e.g.][and references 
therein]{ie95}. The correctness of their classification can thus be tested 
by inspecting their infrared colors. 

The best available infrared sky survey was obtained by the recent Wide-field
Infrared Survey Explorer (WISE, launched in 2010); its all-sky catalog includes
about 560 million objects \citep{WISEpaper1}. WISE mapped the sky at 3.4,
4.6, 12, and 22 $\mu$m with 5-$\sigma$ point source sensitivities better than 
0.08, 0.11, 1 and 6 mJy (corresponding to Vega-based magnitudes 16.5, 15.5, 
11.2 and 7.9, respectively) in unconfused regions on the Ecliptic. The astrometric 
precision for high signal-to-noise sources is better than 0\farcs15. WISE is 
photometrically calibrated to Vega system and thus objects with infrared excess 
should have colors greater than zero (not accounting for the measurement noise). 

\begin{figure}
\epsscale{0.55}
\vskip -0.3in
\plotone{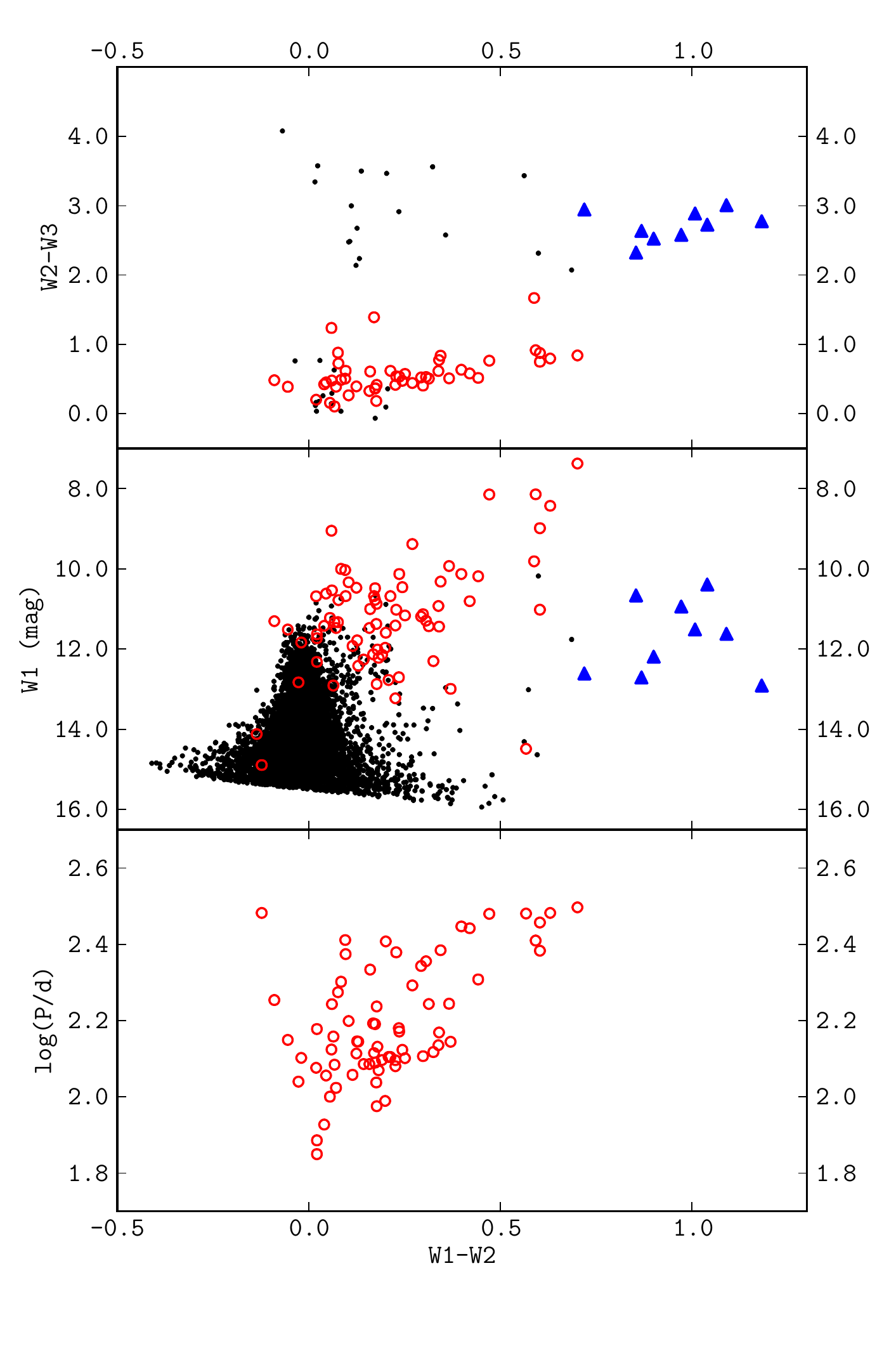}
\vskip -0.6in
\caption{The symbols in the middle panel show the distribution of a subsample of
7,123 variables (out of 7,194) in the PLV catalog that are detected by the WISE
survey and have WISE magnitudes $W1<16.5$ and $W2<15.5$ (5-$\sigma$ detection
limits). Objects classified as ``long-period variables'' (defined as variables
with periods longer than 50 days, and semi-regular variables) are shown as open
circles (74 objects); the majority display infrared excess compared to colors of
dust-free stars ($W1-W2 \sim 0$). Nine objects with light curves classified as
``Other'' and with quasar-like infrared colors, $W1-W2>0.7$, are shown as large
triangles. The top panel shows a WISE color-color diagram for the subset of
99 objects that have $W3<11.2$ (note that the majority of objects without
significant infrared emission do not satisfy this condition; out of 74
long-period variables, 49 satisfy the $W3$ brightness limit, as well as 9
objects with quasar-like colors and 41 other objects). The bottom panel shows the
period-color diagram for the 74 long-period variables. Examples of their light
curves are shown in Figure~\ref{fig:WISElc}. Color version of this figure is
available online. \label{fig:WISE}
}
\end{figure}

We have positionally matched the PLV and WISE catalogs with a matching
radius of 3 arcsec and obtained 7,123 WISE matches for objects listed
in the PLV catalog. Our analysis of this sample is shown in 
Figure~\ref{fig:WISE}. The distribution of WISE colors for objects classified as 
``long-period variables''  is consistent with the majority of them being 
genuine AGB stars \citep{TuWang,Tiss2012}.  Indeed, the brightest and
most famous carbon-rich AGB star, CW Leo (IRC$+$10216) is recovered in our 
sample (LINEAR ID=17154286; $P$=632.511 days based on 475 LINEAR 
measurements; see also \S\ref{sec:resolved}). The paucity of long-period
variables with $W1>13$ is a Galactic structure effect - at high latitudes 
probed by the LINEAR sample (due to the requirement of overlap with the
SDSS footprint) this magnitude cutoff corresponds to several tens of kpc
and thus reaches many disk scale heights away from the plane (Hunt-Walker
et al., in prep.). 

\begin{figure}
\epsscale{0.75}
\vskip -1.3in
\plotone{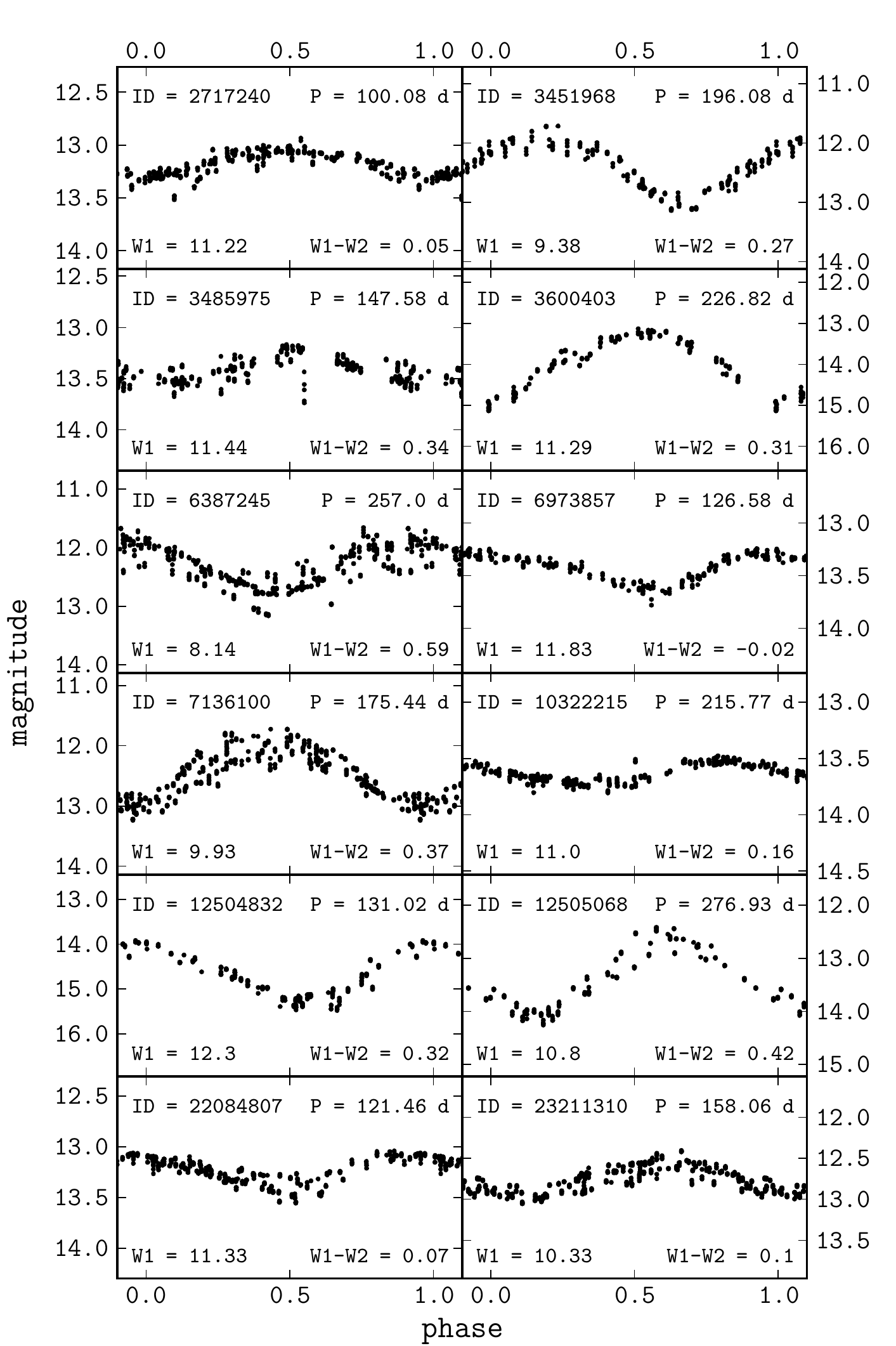}
\vskip -0.2in
\caption{Examples of light curves for objects classified as long-period
variables. Each panel lists LINEAR ID, best-fit period in days, and the WISE
$W1$ magnitude and $W1-W2$ color (see Figure~\ref{fig:WISE}). The scatter in
light curves is much larger than photometric errors and reflects the fact that
light curves for these stars are not exactly reproducible between different
cycles. Color version of this figure is available online. \label{fig:WISElc}
}
\end{figure}

The top panel in Figure~\ref{fig:WISE} shows the period-color relation
for long-period variables. Although there is some correlation between 
the quantities, the scatter is substantial.  The observed scatter in 
$logP$ at a fixed color of about 0.2 dex is in good agreement with 
earlier work \citep[e.g. see][and references therein]{whitelock06}. 
Examples of LINEAR light curves for long-period variables are shown 
in Figure~\ref{fig:WISElc}.  We note that the scatter in phased light curves is 
much larger than photometric errors and reflects the fact that light curves for
these stars are not exactly reproducible between different cycles. 

There are nine objects with light curves classified as ``Other'' that show
infrared colors consistent with quasars \citep[$W1-W2>0.7$, see e.g.][]{yun2012}. 
In addition, there are 14 objects with $W2-W3>2.0$, implying strong infrared 
excess that is likely inconsistent with AGB stars, but also with blue $W1-W2<0.5$ 
colors inconsistent with quasars (Nikutta et al., in prep.). A few but not all of them 
could be chance positional coincidences with background quasars which would 
mostly affect $W3$ and $W4$ measurements (based on a quasar surface density of 
several hundred per square degree and a matching radius of 3 arcsec).

\subsection{Noteworthy objects \label{sec:noteworthy}} 

There are six interesting sources that deserve direct mention by name. There is
one case of a likely type Ia supernova (LINEAR ID=7682813, see the bottom left
panel in Figure~\ref{fig:4nw}) which increased in brightness by 0.8 mag over
about 10 days, and then gradually returned to the initial brightness over about
90 days. The corresponding SDSS image clearly shows a positionally coincident
blue emission-line galaxy at a redshift of 0.028. For the standard cosmology,
the implied absolute magnitude at maximum light is $M = -19.4$, which is
consistent with supernova classification. The absolute magnitude of its blue
host galaxy is $M = -18.6$, in agreement with expectations. The object with
LINEAR ID=17655724 (see the bottom right panel in Figure~\ref{fig:4nw}) steadily
increased in brightness by 0.5 mag over about 5 years. If this trend continues,
in 400 years it would outshine the Sun; nevertheless, this is unlikely because
its SDSS spectrum confirms that this object is a quasar at a redshift of 0.531
\citep[we note that this variability behavior is a bit unusual when compared to
typical quasar variability properties, see e.g.][]{macleod12}. In addition, the
Catalina Sky Survey \citep{dra09} data demonstrate that the brightness increase
is slowing down. 

Given its light curve that shows large variations 
(e.g. a decrease in brightness of $\sim$1 mag over $\sim$200 days; see the top left panel in 
Figure~\ref{fig:4nw}), and its WISE colors, the object with LINEAR ID=2752114 is a good candidate 
for an R Coron\ae{} Borealis star, a supergiant carbon-rich star with episodic mass 
loss \citep{Tiss2012,Tiss2013}. On the other hand, an object with a similar light curve and
WISE colors, LINEAR ID=3766947, is a confirmed BL Lac object at a redshift of 0.1325. 
The object LINEAR ID=7455728 (see the top right panel in Figure~\ref{fig:4nw})   is classified 
as an Algol (EA); it displays a flat-bottom 
primary minimum and frequent faint outliers. While these outliers could be due to
the effects of a nearby (6 arcsec) star, it is not obvious what is the origin of its
very red WISE colors ($W2-W3=2.58$).  Possibly the most curious case is an
optically resolved (see the next section) and spectroscopically confirmed quasar at a redshift 
of 0.152, with quasar-like WISE colors, but with an apparently periodic light curve
(LINEAR ID=23417507, $P \sim 604$ d, amplitude $\sim$ 0.4 mag; see the bottom right panel
in Figure~\ref{fig:SDSS_resolved}). Periodogram of this object shows a strong peak, however the shape of the light curve is not fully repeatable. 
A periodic quasar light curve might have interesting astrophysical implications and
searches for such objects have been reported in the literature. In the largest such
search, \cite{macleod10} found 66 candidates in a sample of $\sim$9,000 quasars
from the SDSS Stripe 82 region with spectroscopic confirmation and SDSS light curves.
They declared them all as unconvincing cases of periodicity because their best-fit periods are 
roughly the same as the span of observations -- that is, only a single putative oscillation
was detected. In contrast, our object displays three full oscillations in the
LINEAR light curve and may be worthy of a followup study. 

\subsubsection{Optically-resolved periodically-variable objects \label{sec:resolved}}

\begin{figure}
\epsscale{0.9}
\vskip -2in
\plotone{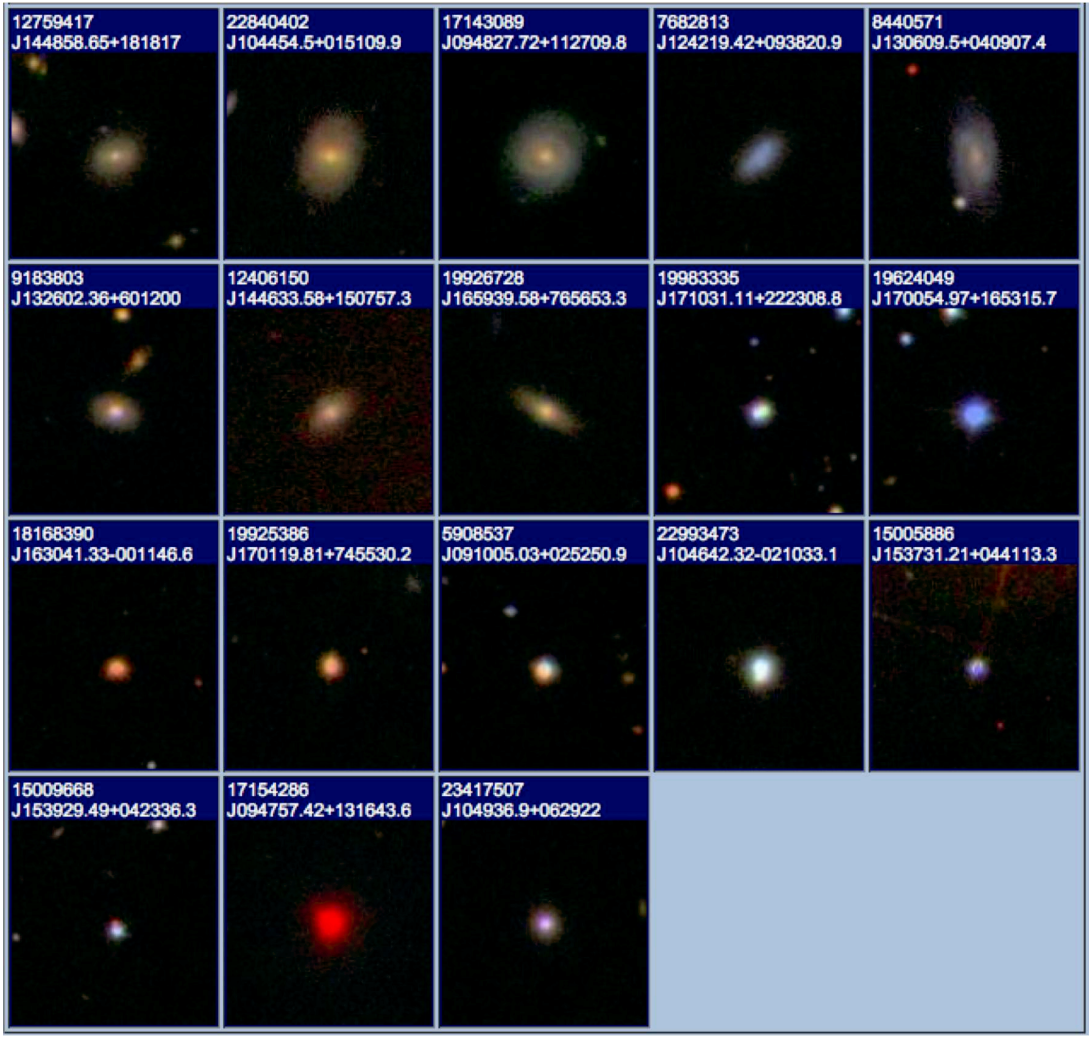}
\vskip -1in
\caption{Default SDSS $gri$ composite images of 18 resolved objects that display
visually confirmed variability in LINEAR data. The top number in each panel is
the object's LINEAR ID. The first eight objects are clearly galaxies. The light
curves for the remaining ten objects are shown in
Figure~\ref{fig:SDSS_resolved}. The extremely red source (the second panel in
the bottom row) is the brightest carbon-rich AGB star, CW Leo (IRC$+$10216).
Color version of this figure is available online. \label{fig:resolvedSDSSimages}
}
\end{figure}

\begin{figure}
\epsscale{0.7}
\vskip -0.3in
\plotone{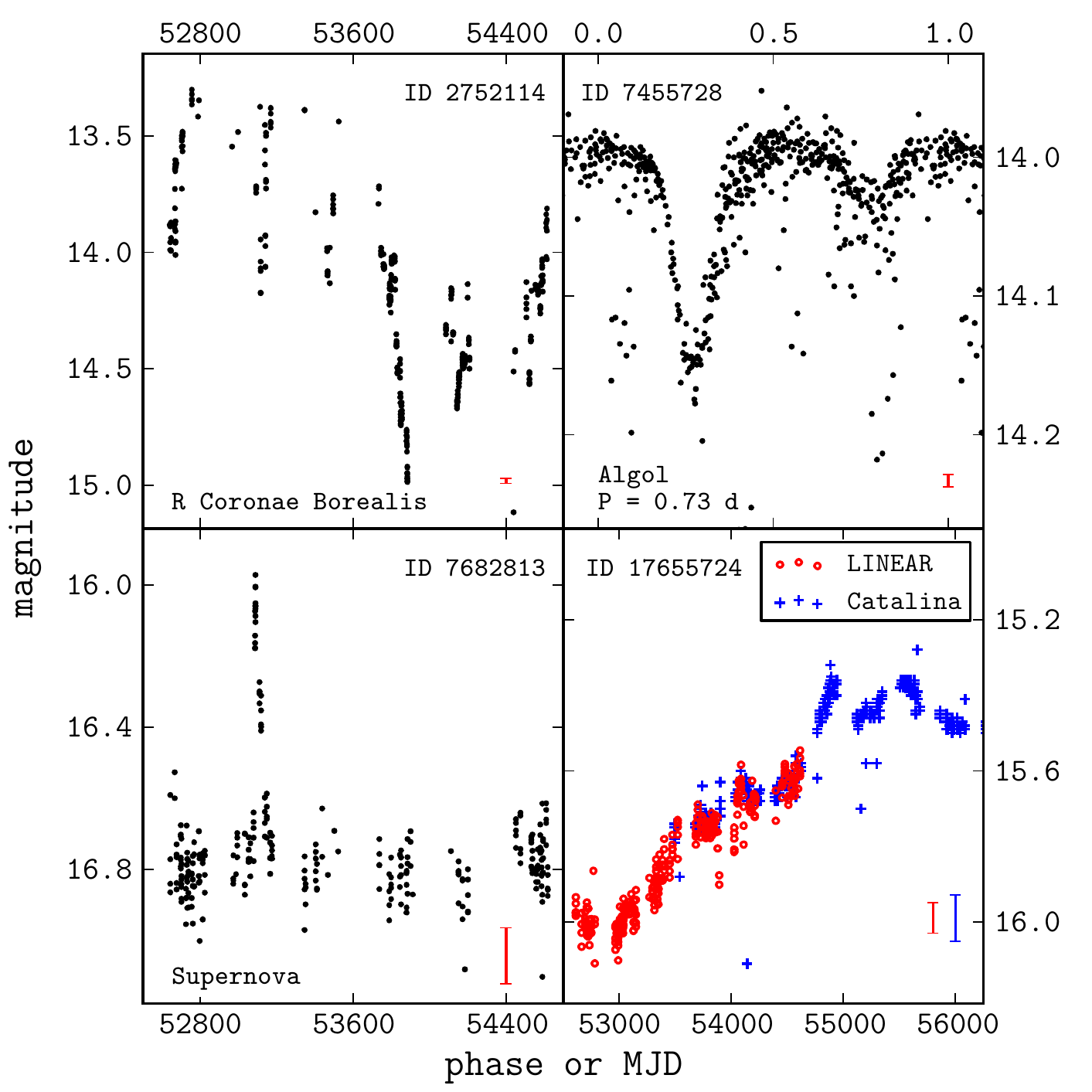}
\vskip -0.2in
\caption{The LINEAR light curves for four objects with unusual light curves (top
left: an R Coron\ae{} Borealis candidate; top right: an Algol-like variable;
bottom left: a supernova candidate; bottom right: a quasar with steady
brightness increase; for more details see \S\ref{sec:noteworthy}). Each panel
lists LINEAR ID and its visual light curve classification from the PLV catalog. 
The vertical error bars show typical photometric errors for each light curve.
The top right panel shows a phased light curve. The bottom right panel also
shows the Catalina Sky Survey data as well (small crosses). Color version of
this figure is available online. \label{fig:4nw}
}
\end{figure}

Among the 7,194 objects listed in the PLV catalog, 18 are optically resolved
(sufficiently large difference between PSF and model magnitudes) in the SDSS
imaging data, and additional 116 objects have unreliable size measurements.
Their SDSS image stamps are shown in Figure~\ref{fig:resolvedSDSSimages}. As
evident, eight objects are clearly galaxies and their variability may be at
least to some extent due to photometric measurement difficulties when using
LINEAR images. Nevertheless, three objects (LINEAR IDs=7682813, 8440571,
9183803) show spectroscopic evidence for AGN activity and their variability may
be real (the last object is also listed in the X-ray ROSAT catalog). 

\begin{figure}
\epsscale{0.7}
\vskip -0.3in
\plotone{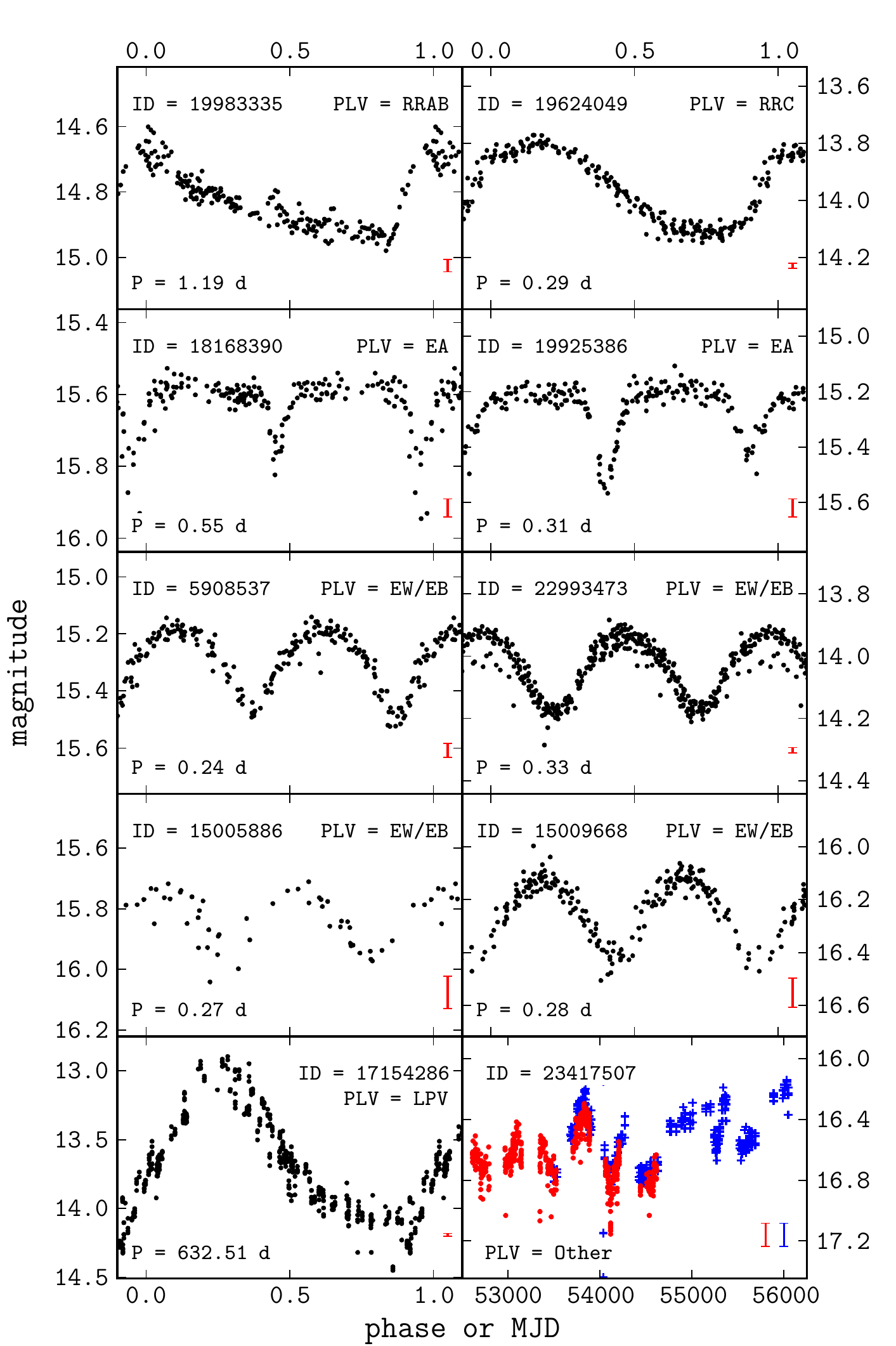}
\vskip -0.2in
\caption{The LINEAR light curves for 10 objects that are optically resolved in
the SDSS imaging data but do not appear as well-resolved  galaxies. Each panel
lists LINEAR ID, its visual light curve classification from the PLV catalog, and
the best-fit period (in days). The vertical error bars show typical photometric
errors for each light curve. All panels except the bottom right panel display
phased light curves. ``The light curve of a quasar in the bottom right panel
combines LINEAR (circles, red in the online version) and CSDR2 (crosses, blue in
the online version) data and confirms its quasi-periodic behavior (note that its
full light curve, and not phased light curve, is shown in this panel).'' Color version of
this figure is available online.
\label{fig:SDSS_resolved}
}
\end{figure}

The light curves for the ten objects that do not appear as well-resolved galaxies are shown 
in Figure~\ref{fig:SDSS_resolved}. Object in the middle right panel (LINEAR ID=22993473,
the fourth object in the third row in Figure~\ref{fig:resolvedSDSSimages}) is beyond doubt
a barely resolved binary system, with a light curve classified as EW/EB. A few sources
show color gradients in their SDSS point spread function (including a known RR Lyr\ae{}
star V368 Her, shown in the top left panel); such gradients can be a sign of their binary 
nature, or possibly of fast changes in the point spread function that led to their misclassification
as resolved objects by the SDSS image processing pipeline \citep{lup02}. The objects shown in the 
bottom row in Figure~\ref{fig:SDSS_resolved} have already been discussed: carbon-rich AGB 
star CW Leo and a quasar with nearly-periodic light curve. For the latter, we have added data 
from the Catalina Sky Survey; during the overlap with the LINEAR data, the two light curves are
consistent. These additional data provide further support for quasi-periodic light variations 
displayed by this quasar.

\section{Classification Based on Machine Learning Algorithms}
 \label{sec:automClass}

We have demonstrated in the preceding section that the distribution of visually-selected 
periodic variables displays distinctive features in the multi-dimensional attribute space 
spanned by the light-curve parameters (period, amplitude, skewness) and optical/infrared 
colors. In this section we explore to what extent can this behavior enable robust and efficient 
automated classification of objects into various classes of variable population.  We consider 
two classification methods based on machine learning algorithms. 

First, we analyze the performance of an unsupervised classification algorithm that attempts 
to recognize existing variability classes in the PLV catalog using only their clustering in the 
multi-dimensional attribute space, but not the results of the visual light curve classification.  
The motivation here is that these clusters correspond to different physical classes of object 
(different types of variable stars) and automated method might pick additional clusters.
We also perform the so-called supervised classification where a training sample is used to 
define selection boundaries. The main goal is to quantify whether visual classification could 
be improved, or perhaps entirely bypassed. 

In order to avoid the impact of objects with unreliable measurements, the starting sample of
7,194 variables is cleaned from sources with unreliable periods, bad SDSS photometry and 
sources without 2MASS detections. We consider only the five most populous classes 
(ab type and c~type~RR~Lyr\ae{},  EA and EW/EB eclipsing binaries and SX~Phoenicis/$\delta$~Scuti
candidates). The resulting cleaned sample of 6,146 variables is publicly available from the
same site as the main 
catalog\footnote{Available from http://www.astro.washington.edu/users/ivezic/r\_datadepot.html}.

\subsection{Unsupervised classification based on a Gaussian Mixture Model \label{sec:GMM}}

The strong clustering of objects, visually classified in six different types using their light curves,
in the multi-dimensional attribute space suggests that an automated unsupervised classification 
scheme might be at least as successful as visual classification (and definitely easier!). 
To investigate this possibility, we used a machine learning algorithm based on a Gaussian 
mixture model to describe the observed distribution of objects. We note that the only
attribute describing light curve shape is skewness. More sophisticated schemes,
such as those based on best-fit parameters for a multi-harmonic Fourier series
fit to light curve, are also possible (e.g., Debosscher et al. 2007; Richards et
al. 2011; and references therein).

\begin{figure}
   \centering
  \vskip -1.5in
   \includegraphics[width=\textwidth]{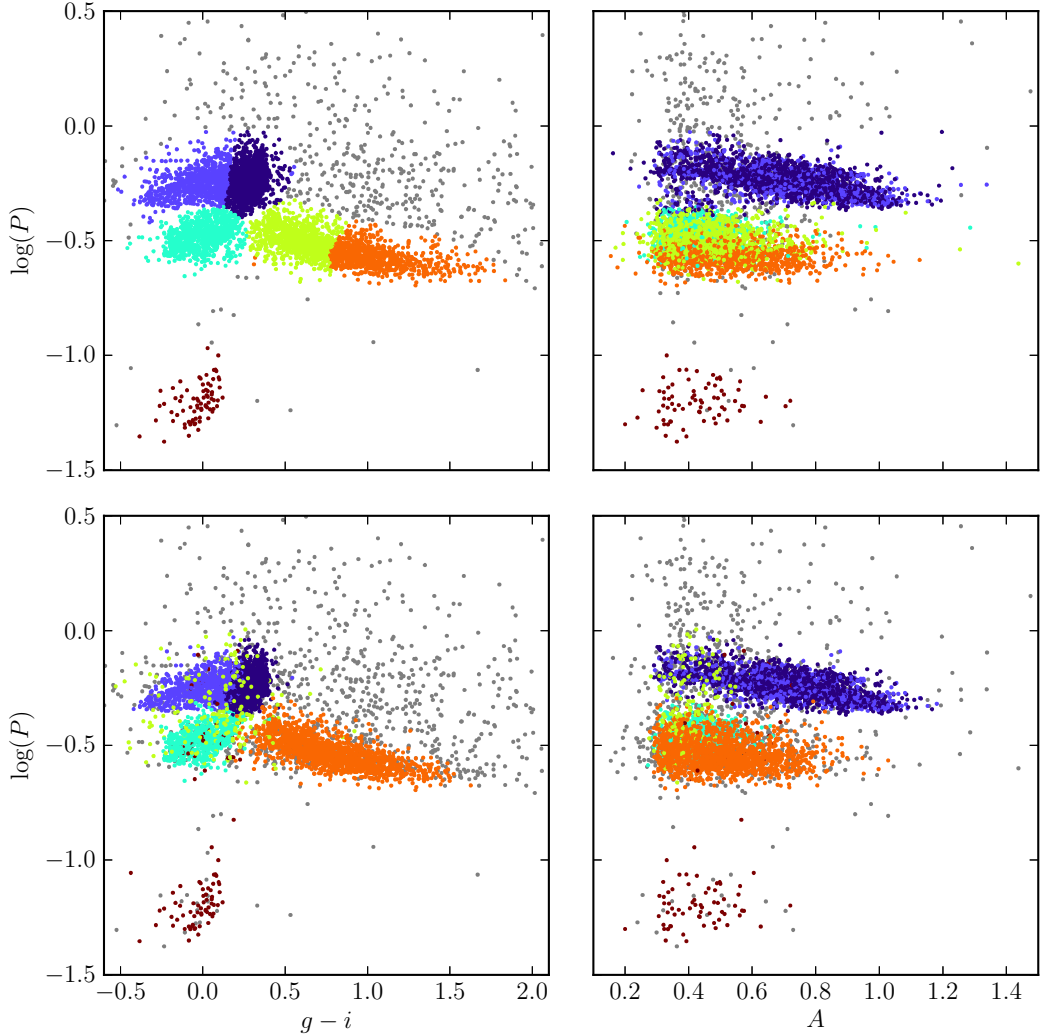}
  \vskip -1.5in
\caption{Unsupervised clustering analysis of periodic variable stars from the
LINEAR dataset using astroML code for Gaussian Mixture Model algorithm.  The top
row shows clusters derived using two attributes ($g-i$ and $\log(P)$) and a
mixture of 12 Gaussians. The colorized symbols mark the six most compact
clusters. The bottom row shows analogous diagrams for clustering based on seven
attributes (colors $u-g$, $g-i$, $i-K$, and $J-K$, $\log(P)$, light curve
amplitude, and light curve skewness),  and a mixture of 15 Gaussians.  See
Figure~\ref{fig:LINEAR_clustering_2} for data projections in the space of other
attributes for the latter case. This figure is adapted from \citet{ive13}, and
can be reproduced using code available at http://www.astroML.org \citep{van12}.
Color version of this figure is available online.
\label{fig:LINEAR_clustering_1}
}
\end{figure}

\begin{figure}
 \centering
\vskip -1.5in
 \includegraphics[width=\textwidth]{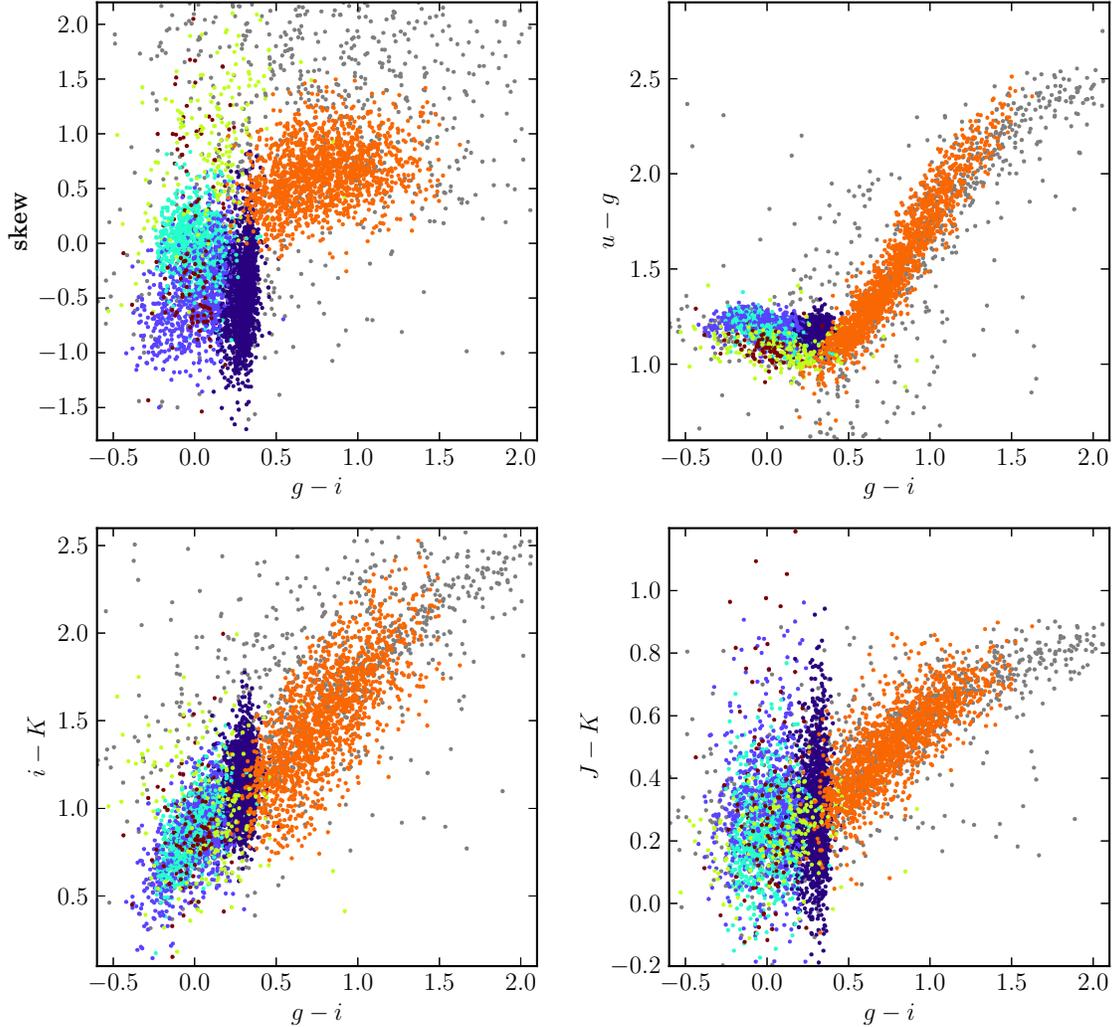}
\vskip -1.5in
\caption{Unsupervised clustering of periodic variable stars from the LINEAR
dataset using Gaussian Mixture Model algorithm. Clusters are derived using seven
attributes (colors $u-g$, $g-i$, $i-K$, and $J-K$, $\log(P)$, light curve
amplitude, and light curve skewness), and a mixture of 15 Gaussians.  The
colorized symbols mark the six most compact clusters. The $\log(P)$ vs. $g-i$
diagram and $\log(P)$ vs. light curve amplitude diagram for the same clusters
are shown in the lower panels of Figure~\ref{fig:LINEAR_clustering_1}. This
figure is adapted from \citet{ive13}, and can be reproduced using code available
at http://www.astroML.org \citep{van12}. Color version of this figure is
available online. \label{fig:LINEAR_clustering_2}
}
\end{figure}

The Gaussian Mixture model (GMM) describes the density of data points using a sum of multi-variate 
Gaussians. Statistically significant clusters of points are assigned a Gaussian, and in case of 
complex cluster morphology, multiple Gaussians. This clustering method does not require a 
training sample and thus belongs to the class of unsupervised classification (clustering) methods. 
The number of required clusters and their best-fit parameters are typically obtained using the
Expectation Maximization method \citep{dem77}. We used a GMM implementation from
{\it astroML}, a set of publicly available\footnote{See http://www.astroML.org} \citep{van12}
data mining and machine learning tools implemented in {\it python}. 
Figures~\ref{fig:LINEAR_clustering_1} and \ref{fig:LINEAR_clustering_2} show the GMM results 
for two cases. 

The top panel in Figure~\ref{fig:LINEAR_clustering_1} shows a 12-component Gaussian mixture 
model using only two most discriminative data attributes, the $g-i$ color and $\log(P)$. 
The number of components is determined automatically using the Bayesian Information
Criterion (see {\it astroML} documentation for details).  Out of the 12 clusters, six are very compact, 
while the rest seem to describe the background. Three clusters correspond to ab and c~type~RR~Lyr\ae{} 
stars. Interestingly, the former are separated into two clusters. The reason is that
the $g-i$ color is a single-epoch color from SDSS that corresponds to a random phase. 
Since ab~type~RR~Lyr\ae{} stars spend more time close to minimum than to maximum light, 
when their colors are red compared to colors at maximum light, their color distribution 
deviates strongly from a Gaussian.  The elongated sequence populated by
various types of eclipsing binary stars is also split into two clusters because 
its shape cannot be described by a single Gaussian either.  The upper-right panel shows the 
clusters in a different projection, $\log P$ vs.~light curve amplitude.  The top four clusters are 
still fairly well localized in this projection due to $\log P$ carrying significant discriminative power.

In another instance of GMM analysis, the clustering attributes included four photometric colors based on 
SDSS and 2MASS measurements ($u-g$, $g-i$, $i-K$, $J-K$) and three parameters determined 
from the LINEAR light curve data ($\log(P)$, amplitude, and light curve skewness). A
15-component Gaussian mixture model to this seven-dimensional dataset yields the clusters
shown in the bottom panels of Figure~\ref{fig:LINEAR_clustering_1}.  The clusters
derived from all seven features are remarkably similar to the clusters derived
from just two features: this shows that the additional data adds very little new
information (equivalently, this shows that the seven attributes are strongly correlated).
The main difference compared to the two-attribute case is that the EB/EW sequence is
now described by a single component. Figure~\ref{fig:LINEAR_clustering_2} shows the locations 
of the six most compact clusters in the space of other attributes. 

As is evident from visual inspection of Figures~\ref{fig:LINEAR_clustering_1} and
\ref{fig:LINEAR_clustering_2}, the most discriminative attribute is the period. A few
clusters which have very similar period distributions, are separated by the
$g-i$ and $i-K$ colors, which are a measure of the star's effective temperature; 
see \cite{cov07}. In summary, although there are many Gaussian components
in the chosen mixture models, no new compact classes were revealed by this 
automated analysis.

\subsection{Supervised classification with Support Vector Machine \label{sec:SVM}}

Given the results of visual classification, we attempt to reproduce it in automated
fashion using supervised classification and a machine learning method called
Support Vector Machine \citep[SVM;][]{CCVV95}.  SVM uses linear classification
boundaries, but unlike our simple method described in \S\ref{sec:boundaries}, 
they do not need to be aligned with the coordinate axes. The optimal classification
boundaries are those that maximize the class separation, or margin (the training
points that are found on the margin are called support vectors). 

\begin{figure}
   \centering
  \vskip -1.5in
   \includegraphics[width=\textwidth]{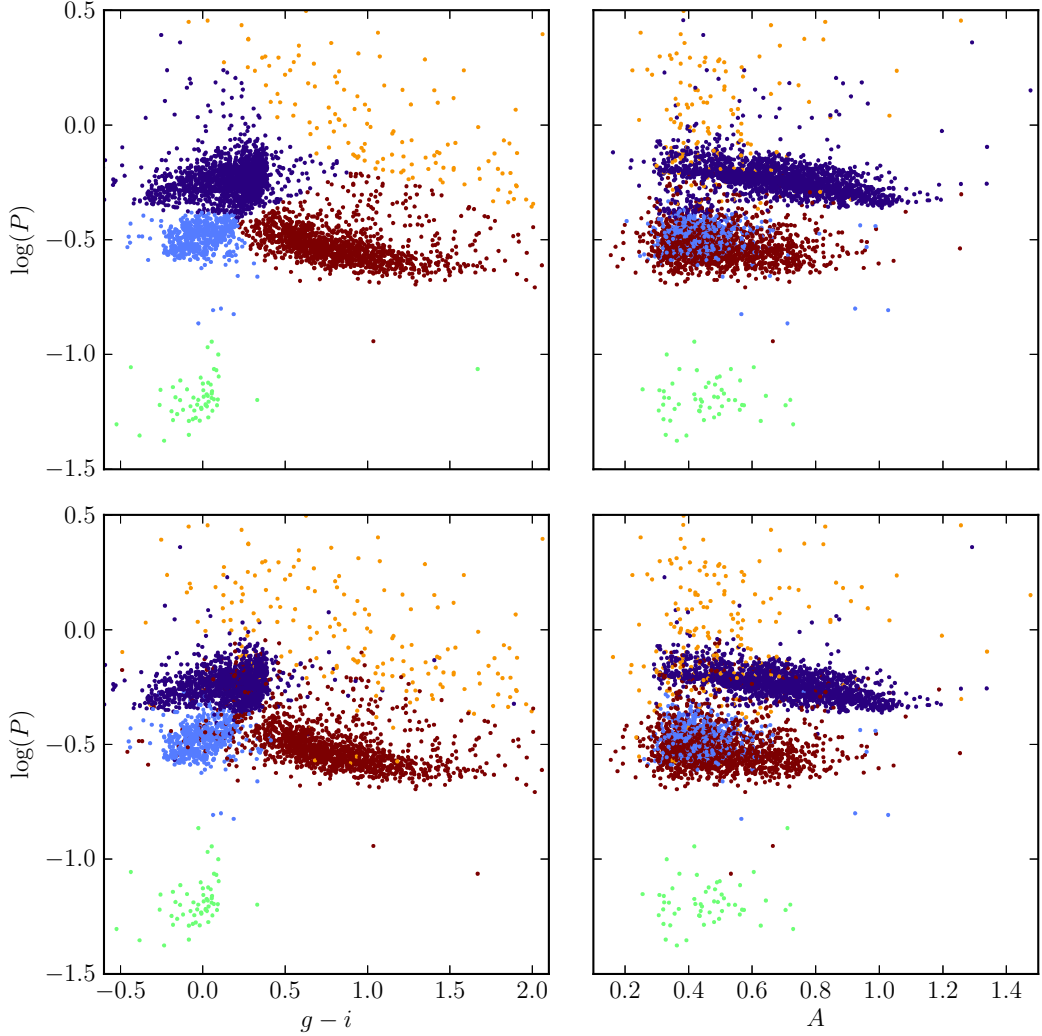}
  \vskip -1.5in
\caption{Supervised classification analysis of periodic variable stars from the
LINEAR dataset using astroML code for Support Vector Machine algorithm.  The top
row shows classes derived using visual classification results for five classes
and two attributes ($g-i$ and $\log(P)$). One third of the sample was used as
training sample. The colorized symbols mark objects from the five classes
adopted by SVM. The bottom row shows analogous diagrams for classification based
on seven attributes (colors $u-g$, $g-i$, $i-K$, and $J-K$, $\log(P)$, light
curve amplitude, and light curve skewness).  See
Figure~\ref{fig:LINEAR_clustering_4} for data projections in the space of other
attributes for the latter case. This figure can be reproduced using code
available at http://www.astroML.org \citep{van12}. Color version of this figure
is available online. \label{fig:LINEAR_clustering_3}
}
\end{figure}

\begin{figure}
 \centering
\vskip -1.5in
 \includegraphics[width=\textwidth]{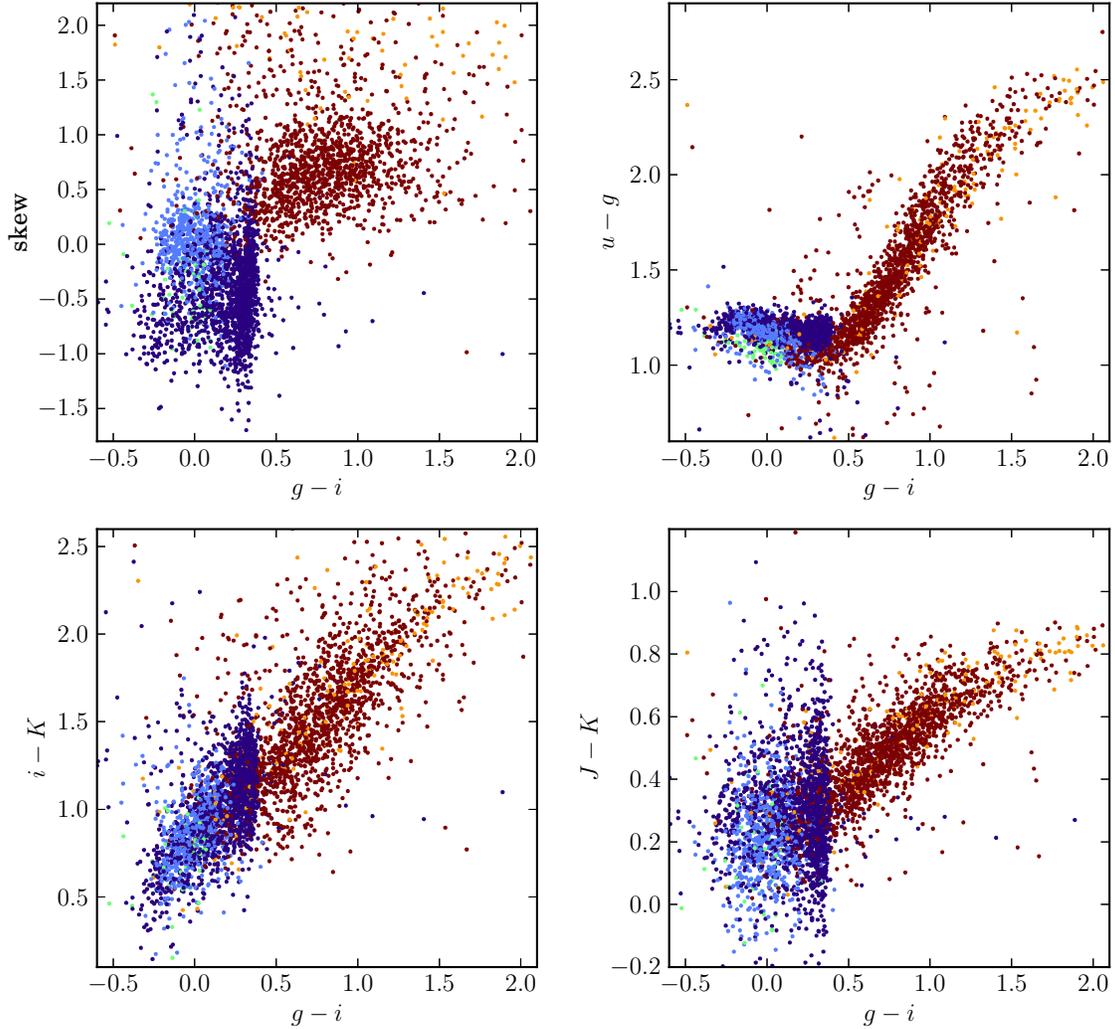}
\vskip -1.5in
 \caption{Supervised classification analysis of periodic variable stars from the
LINEAR dataset using Support Vector Machine algorithm. Classes are derived using
seven attributes (colors $u-g$, $g-i$, $i-K$, and $J-K$, $\log(P)$, light curve
amplitude, and light curve skewness). The colorized symbols mark objects from
the five classes adopted by SVM. The $\log(P)$ vs. $g-i$ diagram and $\log(P)$
vs. light curve amplitude diagram for the same classes are shown in the lower
panels of Figure~\ref{fig:LINEAR_clustering_3}. This figure can be reproduced
using code available at http://www.astroML.org \citep{van12}. Color version of
this figure is available online. \label{fig:LINEAR_clustering_4}
}
\end{figure}

We used a multi-label SVM from the {\it scikit-learn} package \citep{ped11},
via {\it astroML}. A randomly selected third of the sample is used for training
SVM, and the remaining two thirds for measuring the classification performance.
Figures~\ref{fig:LINEAR_clustering_3} and \ref{fig:LINEAR_clustering_4}
illustrate the SVM results for two cases\footnote{This part of analysis can be
easily reproduced using public and open-sourced {\it astroML} code and datasets
available at http://www.astroML.org.}, and Table~\ref{Tab:LINEARclassesSVM}
provides a quantitative summary. 

As with unsupervised GMM clustering, both two-attribute and seven-attribute cases are 
considered. SVM assigns a large fraction of the EA class (Algol-type eclipsing binaries) to
the EB/EW class (contact binaries). This is not necessarily a problem with the SVM method
because these two classes are hard to distinguish given LINEAR light curves. Compared
to the simple method discussed in \S\ref{sec:boundaries}, the precision of SVM classification
relative to visual classification is a bit better (especially for c~type~RR~Lyr\ae{} stars). 
Furthermore, SVM code from {\it astroML} was much easier to deploy than to 
develop the manual method from \S\ref{sec:boundaries}.

\section{Discussion and Conclusions}

We described the creation of a catalog of visually confirmed periodic variable stars
selected from data acquired by the LINEAR asteroid survey, the ``PLV'' catalog. 
The catalog consists of 7,194 variable objects, with over 96\% of entries
that are likely periodic variable stars. Combined with large sky coverage 
($\approx$10,000 deg$^2$) and a flux limit several magnitudes fainter than for most 
other wide angle surveys ($14<r<17$), this catalog can be useful for a wide variety of 
research topics such as studies of Galactic halo structure and the physics of pulsating 
stars and eclipsing binaries. 

The completeness of the PLV catalog, relative to the initial sample of 200,000 candidate
variables, is very high ($>$98\%); nevertheless, it is subject to selection criteria
listed in \S\ref{sec:selection} that were used to select the initial sample subjected to
visual classification. Based on a comparison with the SDSS Stripe 82 variable 
stars, we estimated that the completeness of the PLV catalog is 55--70\%; most of 
the LINEAR incompleteness is due to larger adopted minimum rms variability, 0.1 mag 
vs. 0.05 mag for the SDSS catalog. 

The purity of the PLV catalog is also high as well as the classification
precision ($>$96\% of entries have assigned light curve type). Folded light
curves of all the objects in the catalog were visually inspected several times.
Additional attributes (SDSS, 2MASS and WISE colors) were used to better
characterize each of the objects and thus improve classification purity.
Furthermore, we compared our results to GCVS and VSX variable star catalogs, and
to RR Lyr\ae{} catalogs from the Catalina and Mount Lemmon Surveys (see Appendix
for details) in order to ascertain effectiveness of our method. This analysis
provides further support for the claim of low contamination level by
non-variable objects in the PLV catalog. 

Our analysis was focused on the periodic variables, therefore many irregular and
quasi-periodic variables did not make it into the visual inspection stage or in
case they passed the initial low level statistical cuts were ignored during the
visual classification process. We did, however, stumble upon some of these
non-periodic objects while examining the light curves. Some of those variables
and transients (e.g., active galactic nuclei, AM Herculis, BL Lacertae, BY
Draconis, cataclysmic variables, RS Canum Venaticorum) are grouped in the
``Other'' PLV class. 

This suggests that many other interesting object types could be extracted from
PLV. Many of these are not periodic and therefore we made no true attempt to
classify them. 

The PLV catalog is dominated by RR Lyr\ae{} stars (3,913 or 54\%) and eclipsing binaries
(2,762 or 38\%). We also found 112 (1\%) candidate SX~Phoenicis/$\delta$~Scuti
variables and 77 (1\%) red variables with long regular or semi-regular periods (Mirae,
LPV, SR). As suspected in Introduction, we confirm that variable sources fainter
than $V=14$ are made of quite a different population mix than brighter and
better studied sources. Table \ref{tab:results_VC} describes in detail the content 
of the PLV catalog.

An exciting result of our effort is the discovery of 112 SX~Phe/$\delta$
Sct candidates. It is not possible to differentiate the two on the basis of
light curve attributes and color. However, our preliminary analysis based on
SDSS spectra and radial velocities (see \S\ref{sec:SXPhe} and Figure~\ref{fig:radvelFeH}) shows
that they are consistent with the Population II objects and therefore we assume
that the sample is dominated by SX~Phe stars. Until now these stars have been
found mostly in Galactic globular clusters ($\approx$ 250 objects in total) and 
only 17 field SX~Phe stars are currently known. Therefore, if our assumption is correct, 
the PLV SX~Phe sample would increase the number of currently known such
stars by 30\%, and the number of field SX~Phe stars by as much as a factor of six. This 
increase in the sample size could play an important role in characterizing not only this
type of variables but blue stragglers as well. We are currently undertaking a follow-up 
program using several modest-size photometric telescopes (1.2m and 0.25m). 

We note that SX~Phe/$\delta$~Sct candidates are found in the region of the $u-g$ vs. $g-r$
color-color diagram populated by RR Lyr\ae{} stars, with a number ratio of 1:40. 
Therefore they do not represent a major contaminant of RR Lyr\ae{} samples; 
our results confirm early estimates of the upper limit for their contamination
fraction of 10\% \citep{ive00}. 

Compared to e.g. 10,000 eclipsing binaries in the Galactic bulge fields discovered by OGLE II
and analyzed by \cite{Devor04}, or to $\sim$2,000 eclipsing binaries discovered
in the Kepler survey data \citep{pz05}, our sample of $\sim$2,700 stars is in the sample 
ballpark. Its comparative  advantage is in the large sky area which potentially enables 
studies of the variation of eclipsing binary star properties with location in the Galaxy 
(and by extension, with metallicity and possible other parameters). We note that 
the period distribution for eclipsing binaries in the PLV catalog is generally in 
agreement with previous work, e.g. \citep{Giurcin83,Devor04,pz05}. 

We demonstrated that the availability of SDSS, 2MASS and WISE data can enable analysis 
that is not possible with single-band light curves alone. For example, we derived a precise
quantitative description of an interesting correlation between colors of EB/EW type 
contact binaries and their period (\S\ref{PC_binaries}): as the spectral type (determined 
from $g-i$ SDSS color) of these binaries changes from approximately K4 to F5, their
median period increases form 5.9 to 8.8 hours. Since no consensus about the origin 
of the short-period boundary for contact binaries is reached yet, the improvement in 
observational constraints enabled by LINEAR data will be valuable for future studies 
of stellar evolution. We also showed how WISE colors can be used to better identify
several populations, including asymptotic giant branch stars, R Coron\ae{} Borealis stars and 
quasars. 

We emphasize that the preliminary work described in \S\ref{sec:analysis} is by no means 
a complete analysis of the PLV catalog. To point out but a single example, detailed analysis 
of light curves for eclipsing binaries using more sophisticated methods such as Fourier
analysis, or full physical model fitting \citep{ruc92,Devor04,pz05}, is capable of providing 
valuable further insight into the physics of such stellar systems. In addition, this 
variable stars sample will be valuable for comparison to Gaia results, for example, to 
search for period evolution \citep[e.g.,][]{dav13}. 

We conclude by pointing out that processing the volume of light curve data provided
by the LINEAR survey is still (barely) manageable by human resources. However, with 
the upcoming large surveys, such as Gaia and LSST, automated schemes will have to be
employed to classify the expected vast volumes of data. Examples of such methods, 
based on machine learning algorithms, are discussed in \S\ref{sec:automClass}. In 
addition to the requirement for ever fainter training samples, we point out the need 
for efficient automated recognition of outliers, a problem that we left for the future 
work with the PLV catalog.

\acknowledgments

Authors would like to thank Gaia Coordination Unit 7 (based at ISDC, Department
of Astronomy, University of Geneva, Switzerland) for the help and infrastructure
used in the calculation of Lomb-Scargle and Generalized Lomb-Scargle periods.

L.P. acknowledges support by the Gaia Research for European Astronomy Training
(GREAT-ITN) Marie Curie network, funded through the European Union Seventh
Framework Programme ([FP7/2007-2013] under grant agreement n$\degr$ 264895.
\v{Z}.I. acknowledges support by NSF grants AST-0707901 and AST-1008784 to the
University of Washington, by NSF grant AST-0551161 to LSST for design and
development activity, and by the Croatian National Science Foundation grant
O-1548-2009. The LINEAR program is funded by the National Aeronautics and Space
Administration at MIT Lincoln Laboratory under Air Force Contract
FA8721-05-C-0002. Opinions, interpretations, conclusions and recommendations are
those of the authors and are not necessarily endorsed by the United States
Government.

\appendix

\section{Comparison to Extant Catalogs of Variable Stars}

\subsection{Comparison to General Catalog of Variable Stars and 
                   AAVSO International Variable Star Index}

\begin{figure}
\epsscale{0.5}
\plotone{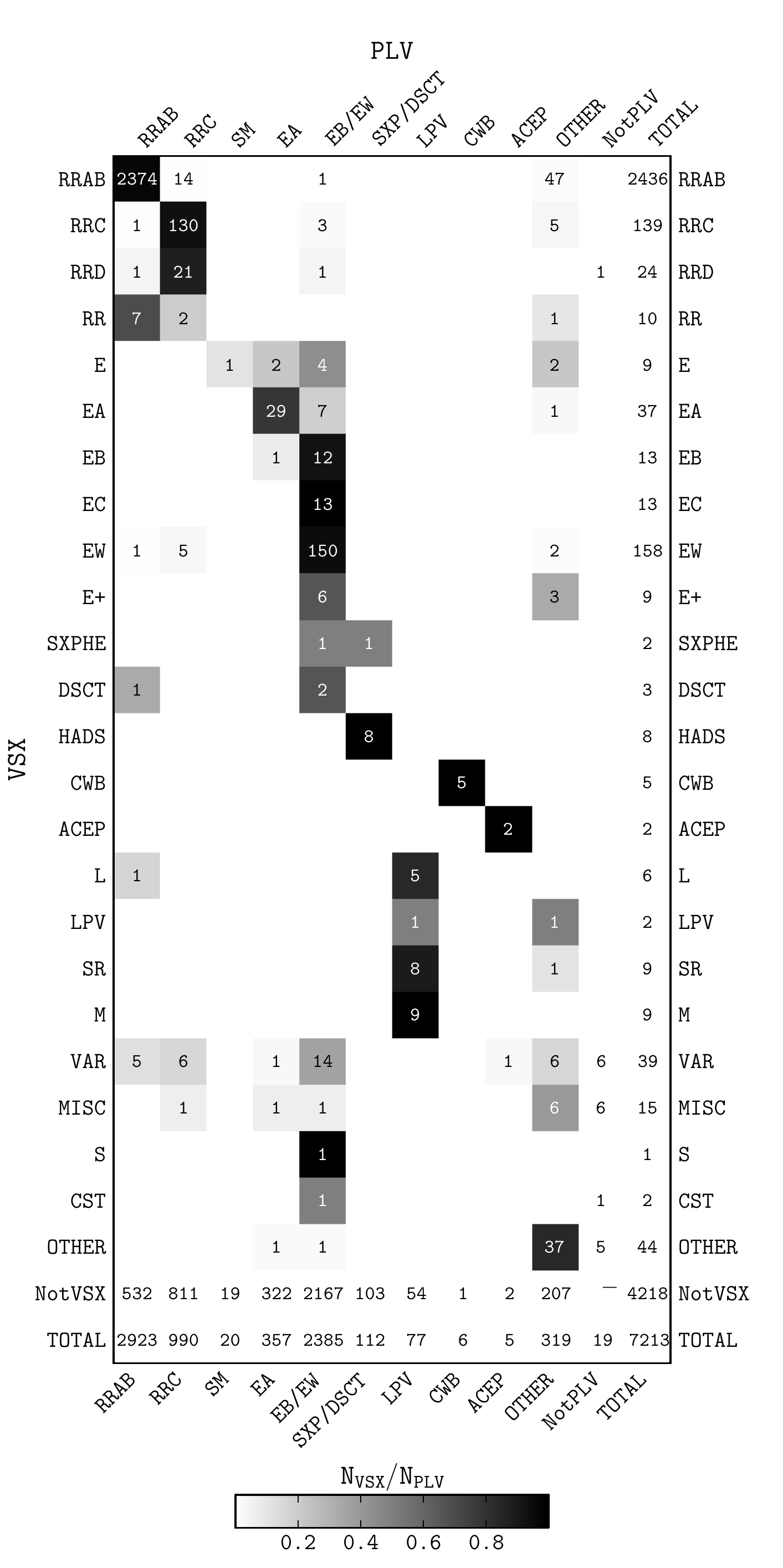}
\caption{The PLV (column) vs. VSX (row) confusion matrix. The column labeled
``Other'' corresponds to variable PLV objects that do not have reliable
variability type. The ``NotPLV'' column corresponds to VSX objects that are not
included in PLV, and the row ``NotVSX'' to PLV objects not listed in the VSX
catalog. The row ``Other'' corresponds to VSX variables with classes others than
listed in this confusion matrix. The intersection regions are color-coded by the
fraction of objects in each row falling into a given region. Acronyms are
according to \citet{VSX2012}. \label{fig:VSX_CM} 
}
\end{figure}

\begin{figure}
\epsscale{0.5}
\plotone{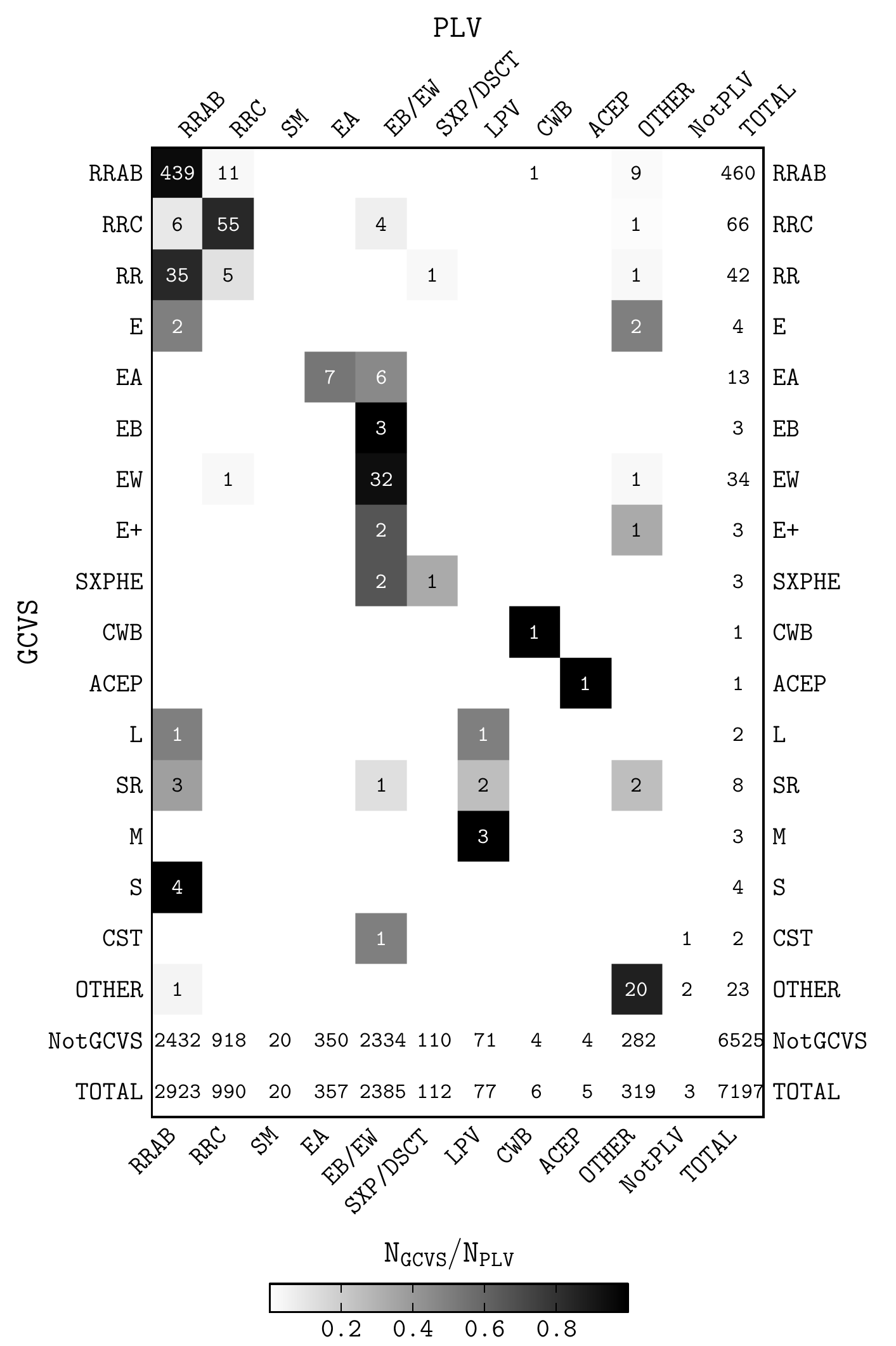}
\caption{The PLV (column) vs. GCVS (row) confusion matrix. The column labeled
``Other'' corresponds to variable PLV objects that do not have reliable
variability type. The ``NotPLV'' column corresponds to GCVS objects that are not
included in PLV, and the row ``NotGCVS'' to PLV objects not listed in the GCVS
catalog. The row ``Other'' corresponds to GCVS variables with classes others
than listed in this confusion matrix. The intersection regions are color-coded
by the fraction of objects in each row falling into a given region. Acronyms are
according to \citet{GCVS2012}. \label{fig:GCVS_CM}
}
\end{figure}

In order to estimate the number of previously unknown variable stars in the PLV
catalog,  we compared it to two online catalogs --- the General Catalog of
Variable Stars \citep[GCVS,][]{GCVS2012} and the American Association of
Variable Star Observers International Variable Star Index \citep[VSX,
][]{VSX2012}.  The Topcat tool \citep{tay05} was used to find positional matches
within 3 arcsec radius (in early February 2013). Our results are summarized in
Figures \ref{fig:VSX_CM} and \ref{fig:GCVS_CM}.

Approximately 60\% of PLV objects could not be matched to an VSX catalog entry,
and approximately 90\% could not be matched to a GCVS entry. We note that the
matching rate for the VSX catalog is higher than for matching to SIMBAD
database: only 1,374 PLV entries, or 19\%, have a SIMBAD object within 3 arcsec
(with 41 different SIMBAD types; they are dominated by RR Lyr\ae{} stars and
non-descriptive ``Star'' types, which account for $\sim$70\% of matches).
Therefore, the majority of PLV entries are previously uncataloged variable
stars. 

\begin{figure}
\epsscale{0.65}
\vskip -0.2in
\plotone{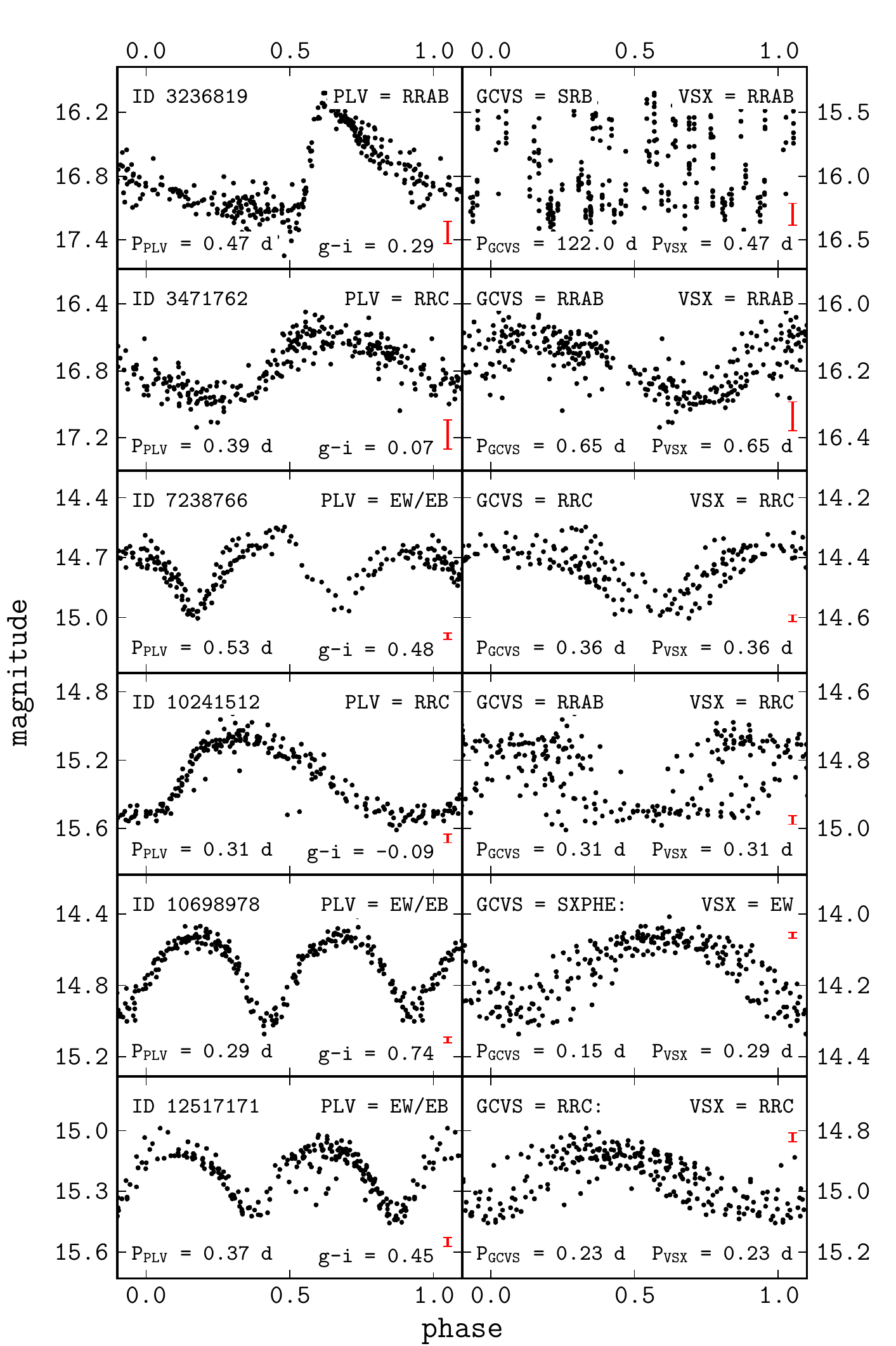}
\vskip -0.1in
\caption{Examples of light curves of objects for which the GCVS or VSX period
and classification does not agree with the PLV classification. The vertical red
bars in each panel in the left column show the median error for LINEAR data.
Plots on the left are folded with the PLV periods, and plots on the right are
folded with the GCVS periods. It is evident that the PLV periods produce
smoother folded light curves and thus are more likely to be correct. The objects
are (top to bottom): BC CVn, GZ Com, V0533 Hya, BE Boo, UW CVn, V0593 Vir.
\label{fig:comp_GCVS_VSX}
}
\end{figure}

For both catalogs, the majority of unmatched objects are eclipsing binaries,
followed by c~type~RR~Lyr\ae{}, SX~Phoenicis/$\delta$~Scuti candidates and long
period variables. Classification of the matched objects shows good overall
agreement between catalogs, and very good agreement for particular types of
objects (e.g. ab~type~RR~Lyr\ae{}). A full visual re-inspection of light curves for
the objects matched in VSX and GCVS was performed, and we stand by our 
classification in all cases. In Figure
\ref{fig:comp_GCVS_VSX} we show several examples where the classification from
GCVS and/or VSX did not match PLV classification.

Comparison to VSX and GCVS motivated us to introduce two more variable star
classes: anomalous Cepheids and BL Herculis. Both can have light curves and
colors that are very similar to those of ab~type~RR~Lyr\ae{}. However, some of
them depart slightly from the locus populated by ab~type~RR~Lyr\ae{} (in the
color-period and other diagrams) and we have adopted VSX and/or GCVS
classification in these cases.

\subsection{Comparison to RR Lyr\ae{} Catalog from the Catalina and Mount Lemmon Surveys}

We also compared our results with the combined RR Lyr\ae{} catalogs assembled by
\citet{dra13a} and \citet{dra13b}. Their Catalina Surveys Data Release 2
(CDSR2)\footnote{Available at http://nesssi.cacr.caltech.edu/DataRelease/}
catalog includes 15,000 ab~type~RR~Lyr\ae{} selected from more than 200 million
light curves obtained by Catalina Schmidt Survey (CSS) and Mount Lemmon Survey
(MLS) over 20,000 deg$^2$ of sky, and to a faint magnitude limit $V=20$. In the
following text we refer to this work as DR13. Approximately 6,460 DR13 objects
are located inside area covered by both PLV and DR13 (approximately $125^\circ <
{\rm R.A.} < 268^\circ$ and $-13^\circ < {\rm Dec} < 65^\circ$). A cut on the
magnitude range that corresponds to brightness of objects potentially included
in PLV ($14 < V < 17$) selects approximately 3,170 ab~type~RR~Lyr\ae{} from
DR13. In further analysis we use these area and magnitude cuts, where
applicable.

\begin{figure}
\epsscale{0.5}
\plotone{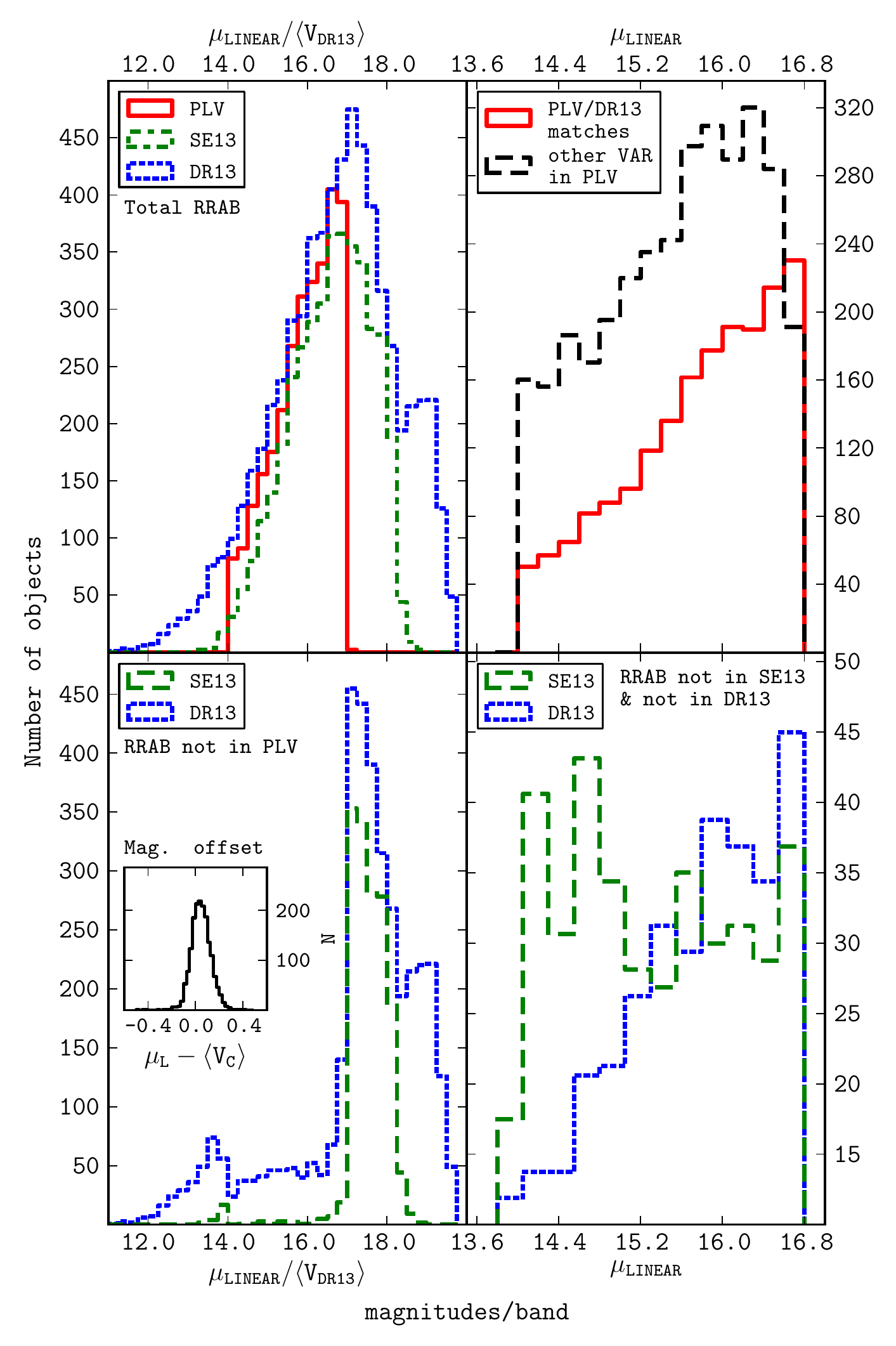}
\caption{Comparison of RR Lyr\ae{} catalogs between PLV, \citet{ses13} and DR13.
The median LINEAR magnitude is designated as $\mu_{LINEAR}$ and mean DR13 V
magnitude as $\left< V_{DR13} \right>$. All four plots are made for the area in
which PLV and DR13 have overlap (approximately $125^\circ <{\rm R.A.} <
268^\circ$ and $-13^\circ < {\rm Dec} < 69^\circ$ ). The histograms in the top
left panel show the number of ab~type~RR~Lyr\ae{} found in these three catalogs.
The histograms in the bottom left panel show ab~type~RR~Lyr\ae{} present in
\citet{ses13} and DR13, but not in the PLV catalog. The inset shows the
difference in brightness for matched objects in the photometric systems used by
LINEAR (unfiltered, $\mu_{L}$) and DR13 (Johnson V band, $\left< V_{C} \right>
$). The histograms in the top right panel show the total number of matched
objects between PLV and DR13, as well as the total number of other variable
stars identified in PLV. The histograms in the bottom right panel show
ab~type~RR~Lyr\ae{} found by PLV and not listed in DR13 (dotted) and
\citet[][dashed]{ses13}. Color version of this figure is available
online.\label{fig:RRAB-histo}
}
\end{figure}

\begin{figure}
\epsscale{0.65}
\vskip -0.2in
\plotone{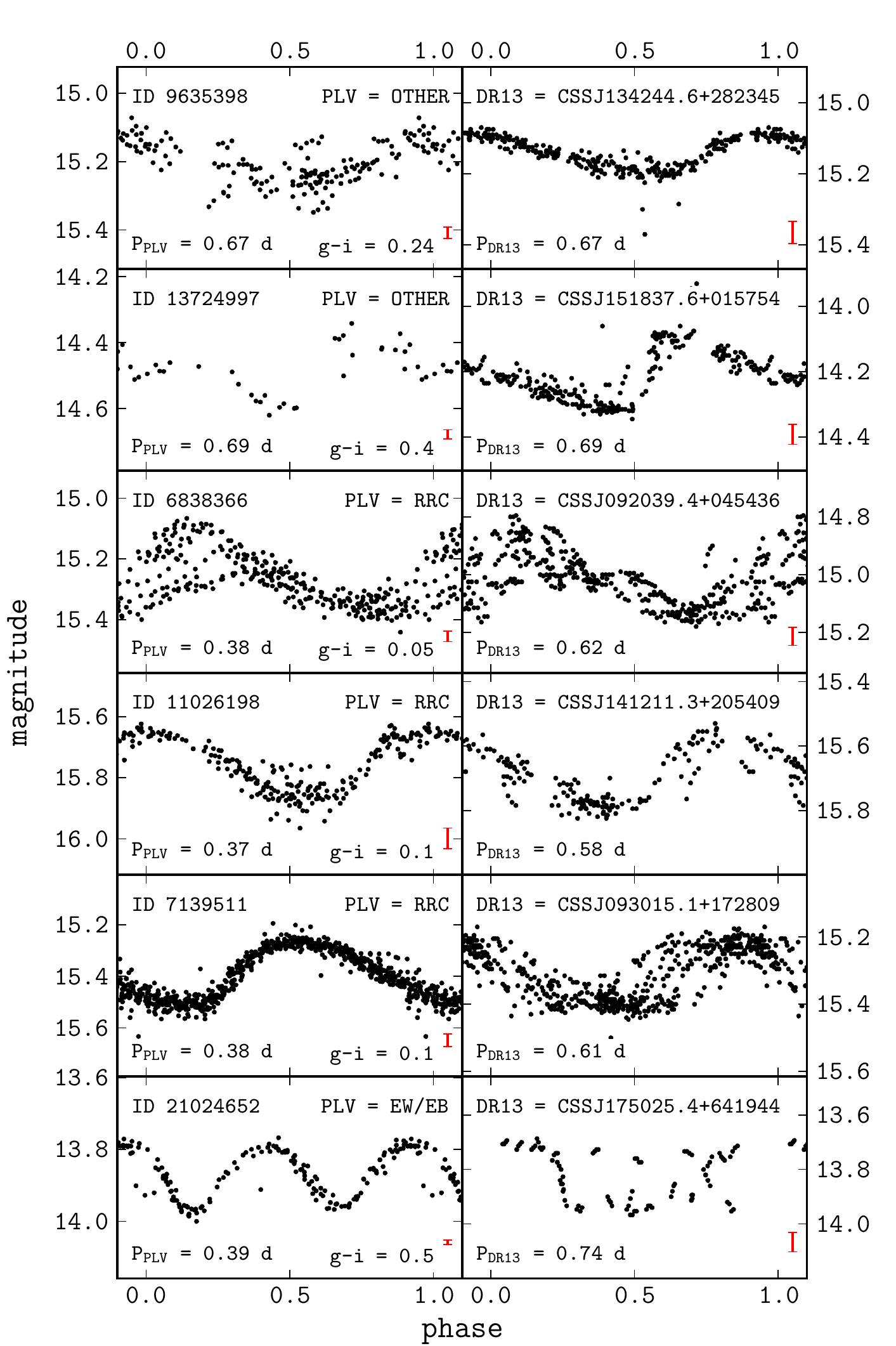}
\vskip -0.1in
\caption{Examples of light curves of objects for which PLV and DR13
classification do not agree. Light curves on the left show LINEAR data folded
with the PLV periods and those on the right show DR13 data folded with the DR13
periods. The vertical error bars in each panel show the median errors (note that
CSDR2 errors are larger than LINEAR errors). All of the objects were classified
as ab type RR Lyr\ae{} in DR13. \label{fig:comp_CRTS}
}
\end{figure}

A 3 arcsec radius match between the initial 200,000 object sample and DR13
selects a total of 2,612 objects (see Figure~\ref{fig:RRAB-histo} for a
statistical summary of the matched sources, which also includes a comparison to
the deeper sample of RR Lyr\ae{} stars from Paper II).  All but 3 are classified
as variable and included in the PLV catalog. Only 86 ($\approx$ 3\%) of the
matched objects are not classified as ab~type~RR~Lyr\ae{} in PLV.  This is a
remarkable agreement level between the two catalogs that were derived from
different datasets and using different techniques. Latter group is dominated by
objects that have poor LINEAR data (66 objects in total) and thus could not be
reliably classified. Their median magnitude and coordinates are distributed
roughly equaly within the PLV brightness range and observed area. These objects
were identified as variable and periodic in PLV, but the light curve type could
not be determined (they are classified as ``Other'' in PLV). Thirteen of the
remaining objects with better data were classified as c~type~RR~Lyr\ae{}, one
was classified as EB/EW eclipsing binary, one as a BL Herculis candidate and two
as anomalous Cepheids (in VSX, these two objects were classified as ACEP and
ACEP:). Therefore, the only true disagreement in classification between LINEAR
and DR13 is for those 13 c~type~RR~Lyr\ae{} (0.5\%). Several examples of light
curves for objects where PLV and DR13 classification did not match are shown in
Figure~\ref{fig:comp_CRTS}. 

\begin{figure}
\epsscale{0.65}
\vskip -0.2in
\plotone{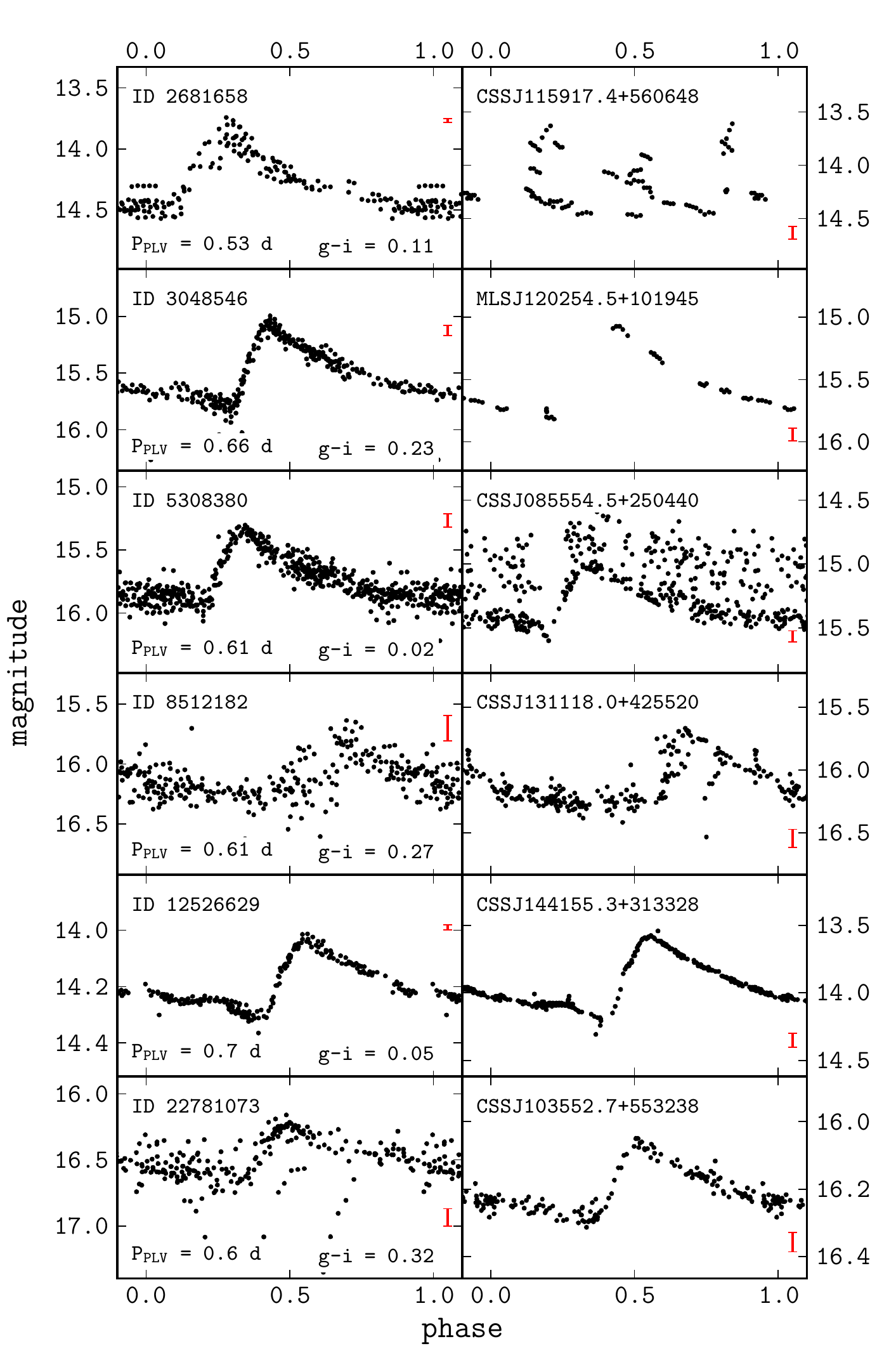}
\vskip -0.1in
\caption{Examples of light curves of ab~type~RR~Lyr\ae{} missing from DR13 but
present in PLV (within the overlapping region). LINEAR data are shown in the
left column and CSDR2 data in the right column. The PLV periods have been used
to fold the light curves. The vertical error bars in each panel show the median
errors. \label{fig:DR13_missing}
}
\end{figure}

Finally, we note that a total of 362 PLV ab~type~RR~Lyr\ae{} (from the
overlaping area and brightness range) do not show up in DR13. Some examples of
these objects are shown in the Figure~\ref{fig:DR13_missing}.


\clearpage

\clearpage

\clearpage

\clearpage

\clearpage

\clearpage

\clearpage

\clearpage

\clearpage

\clearpage

\clearpage

\clearpage

\clearpage

\clearpage

\clearpage

\clearpage

\clearpage

\clearpage

\clearpage

\clearpage

\clearpage

\clearpage

\clearpage

\clearpage

\clearpage

\clearpage

\clearpage
\begin{deluxetable}{cccrrrr}
\tabletypesize{\scriptsize}
\tablecaption{The classification statistics for the SDSS Stripe 82 subsample.
Class is defined by the $A1$ range, the mean classification grade among eight
classifiers, specified in the second and third columns. The standard deviation
among the eight classifiers is listed in the fourth column, and the fifth column
lists the number of light curves in each class. The sixth column lists the
median $\chi^2$ per degree of freedom, and the last column is the median robust
$\chi^2$ per degree of freedom (5\% of the most outlying points are excluded
from the computation). \label{tab:tableA1} } 
\tablewidth{0pt}
\tablehead{
\colhead{Class}  &     \colhead{$A1_{min}$}   &  \colhead{$A1_{max}$} & \colhead{$\sigma_{A1}$}  &     \colhead{$N$}    &  \colhead{$<\chi^2_{dof}>$} & \colhead{$<R\chi^2_{dof}>$}   \\                 
}

\startdata
     0     &   0.0    &   1.1    &      0.38    &  7606    &       9.1   &    1.9  \\
     1     &   1.1    &   1.2    &      0.60    &      75    &     35.4   &     5.9  \\
     2     &   1.2    &   1.8    &      0.73    &      46    &     24.8   &     4.7  \\
     3     &   1.8    &   3.0    &      0.15    &    317    &     28.1   &    13.3  \\
\enddata

\end{deluxetable}

\clearpage
\begin{deluxetable}{ccccc}
\tabletypesize{\scriptsize}
\tablecaption{The rectangular boundaries in the period-amplitude-skewness-color
space used for classification. The boundaries were iteratively tuned to maximize
the ratio of correctly selected and classified objects (with respect to the
visual classification). \label{tab:boundaries}}
\tablewidth{0pt}
\tablehead{
\colhead{Type} & \colhead{log(P) [d]} & \colhead{log(A) [mag]} & \colhead{skewness} & \colhead{g-i}
}
\startdata
ab RR Lyr            & $\langle -0.36, -0.05 \rangle$ & $\langle -0.55, 0.05  \rangle$ & $\langle -1.2,0.2 \rangle$        & $\langle -0.42,0.5 \rangle$  \\
c RR Lyr             & $\langle -0.59, -0.36 \rangle$ & $\langle -0.55, -0.15 \rangle$ & $\langle -0.4,0.35 \rangle$       & $\langle -0.20, 0.35 \rangle$   \\ 
single min           & $> -0.6 $			     & $\langle -0.7, 0 \rangle$      & $\langle  0.32, 3.6 \rangle$      & $\langle -0.2, 3 \rangle$  \\ 
Algol                & $> -0.6 $			     & $\langle -0.67, 0.14 \rangle$  & $\langle  1, 3.7 \rangle$         & $\langle -1.2, 3.8 \rangle$    \\ 
$\beta$ Lyr \& W UMa & $\langle -0.67 ,-0.4 \rangle$ & $\langle -0.56, -0.09 \rangle$ & $\langle -0.1, 1.6 (3.2) \rangle$ & $\langle 0.1(0.2), 1.8 \rangle$ \\ 
SX~Phe/$\delta$~Sct  & $\langle -1.38,-1.05 \rangle$  & $\langle -0.63, -0.12 \rangle$ & $\langle -1.0, 0.7 \rangle$     & $\langle -0.5,0.2 \rangle$      \\
\enddata
\end{deluxetable}

\clearpage

\begin{deluxetable}{clcr}
\tabletypesize{\scriptsize}
\tablecaption{The main light-curve classification results. The first column
lists a numerical class name used in the public catalog, and the second column
lists its more descriptive names.  The number of visually confirmed variable
stars, with purity exceeding 99\% (RR Lyr\ae{}, SX~Phoenicis/$\delta$~Scuti),
and 98\% in case of eclipsing binaries, is listed in the fourth column, and the
fraction of all cataloged objects in a given class is listed in the third
column. Class ``SM'' corresponds to flat light curves with a single minimum, and
class ``Other'' contains periodic variables which could not be reliably
classified and non--periodic variables. \label{tab:results_VC}}
\tablewidth{0pt}
\tablehead{
\colhead{Class} & \colhead{Type} & \colhead{F [\%]} & \colhead{N} 
}
\startdata
1	& RRAB		& 41		& 2923		\\
2	& RRC		& 14		& 990		\\
3	& SM		& $<$1		& 20		\\
4	& EA		& 5		& 357		\\
5	& EB/EW		& 33		& 2385		\\
6	& SXP/DSCT	& 2		& 112		\\
7	& LPV		& 1		& 77		\\
8	& Hearbeat	& $<$1		& 1		\\
9	& BL Her		& $<$1		& 6		\\
11	& ACEP		& $<$1		& 5		\\
0	& Other		& 4		& 318		\\
	& Total		& 100		& 7194		\\
\enddata
\end{deluxetable}

\clearpage

\begin{deluxetable}{crrrrrr}
\tabletypesize{\scriptsize}
\tablecaption{The performance of supervised classification using Support Vector
Machines method in the seven-attribute case. Each row corresponds to an input
class listed in the first column (ab RRL: ab~type~RR~Lyr\ae{}; c RRL:
c~type~RR~Lyr\ae{}; EA: Algol-type eclipsing binaries; EB/EW: contact eclipsing
binaries; SX~Phe: SX~Phe and $\delta$~Sct candidates). The second column lists
the number of objects in each input class, and the remaining columns list the
percentage of sources classified into classes listed in the top row. The bottom
row lists classification contamination in percent for each class listed in the
top row. \label{Tab:LINEARclassesSVM}}
\tablewidth{0pt}
\tablehead{
\colhead{Class} & \colhead{N} & \colhead{RRAB} & \colhead{RRC} & \colhead{EA} & \colhead{EB/EW} & \colhead{SX~Phe}
}
\startdata
 RRAB	&  1772  & 95.9  &   0.3 &   1.4  &   2.4  &  0.0  \\
 RRC	&    583  &  1.5   & 91.3 &   0.2  &   7.0  &  0.0  \\
 EA	&   228   &  5.3   &  1.3  & 67.5  & 25.9  &  0.0  \\
 EB/EW	& 1507   &  2.1   &  4.0  &   3.1  & 90.7  &  0.1   \\
 SX~Phe	&     56   &  0.0   &  1.8  &   0.0   &  1.8  &  96.4  \\
\hline
Purity	&   $-$   &  97   & 88.4   & 68.4 &  90.5  & 98.2     \\
\enddata
\end{deluxetable}

\clearpage

\begin{deluxetable}{ccccccccccccccc}
\tabletypesize{\scriptsize}
\rotate
\tablecaption{{\bf PLV catalog: light curve data.}\label{tab:PLV_cat1} Detailed
description of entries is provided in the table header. Only the first entry is
shown for illustration.}
\tablewidth{0pt}
\tablehead{
\colhead{ID} & \colhead{LCtype}   & \colhead{P}     & \colhead{A}       & \colhead{mmed} & 
\colhead{stdev}    & \colhead{rms}   & \colhead{L$\chi^2_{pdf}$} & \colhead{nObs}   & \colhead{skew}   & \colhead{kurt}  
 & \colhead{L$R\chi^2$}  & \colhead{CUF}    & \colhead{t2}    & \colhead{t3} }
\startdata
2522   &     5   &  0.238812   & 0.68   & 17.00  &  0.22  &  0.25 &   0.543  &  225 &   0.75  &  0.11  & 0.317 & 1 &  0 &  0 \\ 
\enddata
\end{deluxetable}

\begin{deluxetable}{cccccccccccccccc}
\tabletypesize{\scriptsize}
\rotate
\tablecaption{{\bf PLV catalog: SDSS data.}\label{tab:PLV_cat2} 
Detailed description of entries is provided in the table header.}
\tablewidth{0pt}
\tablehead{
\colhead{ID} & \colhead{RA} & \colhead{Dec} & \colhead{oType} & \colhead{nS} & \colhead{rExt} & 
\colhead{u} & \colhead{g}    & \colhead{r}  & \colhead{i}  & \colhead{z}   & 
\colhead{uErr} & \colhead{gErr} & \colhead{rErr} & \colhead{iErr} & \colhead{zErr} }
\startdata
2522  & 117.99$\cdotp$  & 48.67$\cdotp$ &  6 & 1 & 0.139 & 20.24 & 18.24 & 17.28 & 16.89 & 16.67 &  0.06 &  0.01 &  0.01 &  0.01  & 0.01 \\
\enddata
\end{deluxetable}

\begin{deluxetable}{ccccccccccccccc}
\tabletypesize{\scriptsize}
\rotate
\tablecaption{{\bf PLV catalog: 2MASS and WISE data.}\label{tab:PLV_cat3} 
Detailed description of entries is provided in the table header.}
\tablewidth{0pt}
\tablehead{
\colhead{ID} & \colhead{J} & \colhead{H} & \colhead{K} & \colhead{Jerr} & \colhead{Herr} & \colhead{Kerr} &
\colhead{W$_1$} & \colhead{W$_2$} & \colhead{W$_3$} & \colhead{W$_4$} &
\colhead{W1err} & \colhead{W2err} & \colhead{W3err} & \colhead{W4err} }
\startdata
2522 &  15.45 &  14.94 &  14.76  &  0.06  &  0.09  &  0.10 &  14.52 &  14.63 &  12.60  &  8.84  &  0.03  &  0.07 & -9.90 & -9.90 \\
\enddata
\end{deluxetable}

\end{document}